\documentclass[%
 reprint,
 amsmath,amssymb,
 aps,
]{revtex4-2}

\usepackage{graphicx}
\usepackage{dcolumn}
\usepackage{bm}

\usepackage[utf8]{inputenc} 
\usepackage[T1]{fontenc}    
\usepackage{booktabs}       
\usepackage{amsfonts}       
\usepackage{nicefrac}       
\usepackage{microtype}      
\usepackage{amsmath,amssymb}
\usepackage{algorithm,algpseudocode}
\usepackage{booktabs}
\usepackage{comment}

\numberwithin{equation}{section}
\numberwithin{figure}{section}
\numberwithin{table}{section}

\newcommand{\Yset}{\mathcal{Y}}
\newcommand{\Q}{\mathbb{Q}}

\newcommand{\Xset}{\mathcal{X}}
\newcommand{\Ypath}{\mathbb{Y}}

\newcommand{\Xpath}{\mathbb{X}}

\newcommand{\forward}{\Prob_{\mathrm{F}}}

\newcommand{\pif}{\pi_{\mathrm{F}}}
\newcommand{\pib}{\pi_{\mathrm{B}}}
\newcommand{\pibm}{\bar{\pi}_{\mathrm{B}}}
\newcommand{\pibAlpha}{\pi_{\learningRate}}
\newcommand{\pibAlphaM}{\bar{\pi}_{\learningRate}}
\newcommand{\suptime}[2]{#1^{(#2)}}
\newcommand{\KL}[2]{\mathcal{D}\left[{#1} \middle \| {#2}\right]}

\newcommand{\tranF}{\mathbb{T}_{\mathrm{F}}}

\newcommand{\fitness}[2]{F_{#1}(#2)}
\newcommand{\scaledFitness}[2]{f_{#1}(#2)}

\newcommand{\learningRule}{\mathcal{L}}
\newcommand{\learningRate}{\alpha}
\newcommand{\timeEstimateRetro}{\tau_{\mathrm{est}}}
\newcommand{\retroEst}{j_{\mathrm{est}}}
\newcommand{\retroEmp}{j_\mathrm{emp}}
\newcommand{\sphere}[2]{D_{#2}({#1})}
\newcommand{\fim}{I_\lambda} 

\newcommand{\Nmax}{N_{\mathrm{max}}}

\newcommand{\popSigma}{\tilde{\Sigma}} 
\newcommand{\varianceTerm}[1]{\suptime{\popSigma}{#1}}
\newcommand{\KLterm}[1]{\suptime{\mathrm{KL}}{#1}}
\newcommand{\expectedGain}[1]{\Delta_{\mathrm{ex}} \suptime{\lambda}{#1} }
\newcommand{\actualGain}[1]{\Delta_{\mathrm{ac}} \suptime{\lambda}{#1}  }

\newcommand{\argmax}{\mathop{\rm argmax}\limits}

\newcommand{\Prob}{\mathbb{P}}
\newcommand{\R}{\mathbb{R}}

\newcommand{\V}{\mathbb{V}}
\newcommand{\cov}[3]{\operatorname{Cov}_{#3}\left[{#1}, {#2}\right]}
\newcommand{\logV}{\operatorname{log-\mathbb{V}}}
\newcommand{\logCov}[3]{\operatorname{log-Cov}_{#3}\left[{#1}, {#2}\right]}
\newcommand{\logCovAlpha}[3]{\operatorname{log-Cov}^{\learningRate}_{#3}\left[{#1}, {#2}\right]}
\newcommand{\average}[2]{\left\langle {#1} \right\rangle_{#2}}

\newcommand{\trace}[1]{\mathrm{Tr}\left({#1}\right)}
\newcommand{\normal}[2]{\mathcal{N}({#1},{#2})}

\newcommand{\eqnref}[1]{(\ref{#1})}

\usepackage{color}
\usepackage{CJKutf8}
\usepackage{ifthen}
\newcommand{\COMM}[2]{{
\begin{CJK}{UTF8}{ipxm}
\ifthenelse{\equal{#1}{SN}}{\color{blue}}{
\ifthenelse{\equal{#1}{TJK}}{\color{red}}{
\ifthenelse{\equal{#1}{AA}}{\color{cyan}}{
\ifthenelse{\equal{#1}{BB}}{\color{magenta}}}}}
[#1: #2]
\end{CJK}
}}

\usepackage{amsthm}
\theoremstyle{definition}

\usepackage[final]{showlabels}

\begin{document}

\preprint{APS/123-QED}

\title{ Acceleration of Evolutionary Processes by Learning \\and Extended Fisher's Fundamental Theorem
}

\author{So Nakashima$^1$}
 \email{so_nakashima@mist.i.u-tokyo.ac.jp}
\author{Tetsuya J. Kobayashi$^{1,2,3}$}%
 \email{tetsuya@mail.crmind.net}
\affiliation{%
$^1$Faculty of Mathematical Informatics, the Graduate School of Information Science and Technology, the University of Tokyo, 7-3-1, Hongo, Bunkyo-ku, Tokyo, 113-8654, Japan,
}%
\affiliation{
$^2$Institution of Industrial Science, the University of Tokyo, 4-6-1, Komaba, Meguro-ku, Tokyo, 153-8505, Japan.
}
\affiliation{
$^3$Universal Biology Institute, The University of Tokyo, 7-3-1, Hongo, Bunkyo-ku, 113-8654, Japan.}

\date{\today}

\begin{abstract}
Abstract:
Natural selection is general and powerful concept not only to explain  evolutionary processes of biological organisms but also to design engineering systems such as genetic algorithms and particle filters.
There is a surge of interest, both from biology and engineering, in  considering natural selection of intellectual agents that can learn individually. 
Learning by individual agents of better behaviors for survival may accelerate the evolutionary processes by natural selection.
We have accumulating pieces of evidence that organisms can transmit its information to the next generation via epigenetic states or memes.
Also, such idea is important for engineering applications to improve the genetic algorithms and the particle filter.
To accelerate the evolutionary process, an agent should change their strategy so that the population fitness increases the most.
Equivalently, an agent should update the strategy towards a gradient (derivative) of the population fitness with respect to the strategy.
However, it has not yet been clarified whether and how an agent can estimate the gradient and accelerate the evolutionary process.
We also lack methodology to quantify the acceleration to understand and predict the impact of learning.
In this paper, we address these problems.
We show that an learning agent can accelerate the evolutionary process by proposing ancestral learning, which uses the information transmitted from the ancestor (ancestral information) via epigenetic states or memes.
Numerical experiments show that ancestral learning actually accelerates the evolutionary process.
We next show that the ancestral information is sufficient to estimate the gradient.
In particular, learning can accelerate the evolutionary process without communications between agents.
Finally, to quantify the acceleration, we extend the Fisher's fundamental theorem (FF-thm) for natural selection to ancestral learning.
The conventional FF-thm relates the speed of evolution by natural selection to the variety of the individual fitness in the population.
Our extended FF-thm relates  the acceleration of the evolutionary process to the variety of individual fitness of the agent.
By the theorem, we can quantitatively understand when and why learning is beneficial.
\end{abstract}

\maketitle


\section{Introduction}
\label{sec:introduction}
A fundamental question in evolutionary biology is how organisms acquire sophisticated traits, functions, and strategies to survive in harsh and ever-changing environments.
Attempts to answer the question have invented the theory of natural selection~\cite{urry2016campbell}.
Evolutionary process by natural selection is general and powerful enough not only to explain various biological phenomena but also to be applied to optimization of engineering systems.
Genetic and evolutionary algorithms~\cite{back1996evolutionary} solve mathematical optimizations by simulating the ``evolution'' of candidates of the solution.
Also, particle filters are designed to solve the filtering problem of latent state models by approximating a posterior distribution with a population of replicating particles~\cite{bishop2006pattern}.

While the original natural selection is a passive process in that the trait of an organism can change only randomly, several studies both in biology and engineering have considered natural selection of intelligent \emph{agents} (Fig.~\ref{fig:model} (a)) that can \emph{learn} from experience and actively change their traits accordingly.
For biological systems, researchers have discussed the fitness value of information processing of organisms like sensing of environments~\cite{haccou1995optimal, kussell2005phenotypic, rivoire2011value, kobayashi2015fluctuation}.
In this context, some studies~\cite{xue2016evolutionary, kobayashi2019fitness} pointed out the possibility that learning can accelerate the evolutionary process by natural selection.
While this idea seems to violate the conventional assumption that the evolution is a blind watchmaker, we have accumulating pieces of evidence that organisms can transmit its information to the next generation not only via genes but also via epigenetic states or memes~\cite{dawkins2016selfish}.
Epigenetic states and memes enable the organism to transmit the information that is necessary for learning.
A pioneering study by Xue and Leibler~\cite{xue2016evolutionary} considered
a growing population of agents, each of which follows a learning rule to choose the same type as that its parent chose more frequently than the parent (we call it Xue's rule).
They showed that this simple learning rule can acquire the optimal type-switching strategy for changing environments.
In another line of works, the effect of learning on evolution has also been discussed as the Baldwin effect~\cite{baldwin1896new, baldwin1897organic}.

In engineering, it has been shown that genetic algorithms and particle filters can be improved by introducing learning by individual agents or particles.
A memetic algorithm~\cite{moscato1989evolution} and an information geometric optimization~\cite{ollivier2017information} are such extended optimization algorithms that employ an active update of candidates of the solutions by, for example, gradient descent.
In addition, some estimation algorithms of a latent state model employ a population of replicating particles that also individually learn the parameters of the model~\cite{Kantas2009overview}.

In this work, we aim to understand the impact of learning in evolutionary processes both qualitatively and quantitatively from a general view point.

\subsection{Learning in evolutionary processes}
Because the interplay of natural selection and learning is tangled, we firstly describe the situation we consider and the definition of learning in this work.

We consider a population of agents that assexually replicate. 
Each agent has type, and stochastically selects one type in one generation. 
The type and the state of environment can affect the number of offspring that the agent can generate. 
For biological systems, the type can be interpreted as a phenotypic trait of an organism. 
The stochastic type selection can be beneficial when the state of environment changes over time.
The type cannot be directly cut indirectly inherent between generations~\cite{seger1987bet, imke2011bet}. 
Each agent also has a type-switching strategy that determine the probability to choose each type. 
We assume that the strategy is heritable and also subject to selection. 
From the biological viewpoint, the strategy can be regarded as a genetic or epigenetic trait, and the types (phenotypes) of agents can be correlated among generations via inheritance of the strategy.
From the engineering viewpoint, the strategy is related to hyper-parameters that determines the behaviors of an agent.

Since the strategy is heritable, better strategies can be selected via natural selection if we have a diversity of strategies in a population.
In a conventional evolutionary process, the diversity of strategies is generated by random (mutational or epignetic) changes that occur when the strategy of individual agent is inherited from one generation to the next. 
As learning of individual agents, we consider here the case that the inherited strategy is biased based on the past information of ancestors or the population.
Specifically, we consider the learning rules that biases the offspring strategy to gain greater fitness.
In general, the conventional random changes can also be regarded as kinds of learning rule in which no average gain of fitness is expected. 
Therefore, we call them passive or zero-th order learning rules.
Our main focus here is the learning rules that can bias the strategy to have an average gain of fitness.
They should update the strategy to the direction called a \emph{gradient} of fitness, into which the fitness increases (defined rigorously in Section~\ref{sec:universal}).
We call them active or first order because the gradient is closely related to the first derivative of the fitness with respect to the strategy~\footnote{If the second or the higher derivative is used, then the learning is called the second (or the higher)-order}.

Under the setting above, we have at least three problems about the interplay between natural selection and learning.
The first one is whether or when learning can accelerate the evolutionary process
of an agent to acquire the optimal strategy.
Since a learning rule must be simple enough to be implemented in biological systems, we should investigate whether the evolutionary process is accelerated even by simple learning rules.
For engineering systems, such simplicity is desirable for building a scalable learning algorithm.
In the previous work by Xue and Leibler~\cite{xue2016evolutionary}, the simple Xue's rule was shown to achieve the optimal strategy under a constant environment via the evolutionary process. 
However, the zero-th order random changes can also achieve the optimal one and therefore the active learning may not always be beneficial nor efficient compared with passive ones.

The second problem is whether an agent can estimate the gradient from accessible information or not.
In particular, we do not know what information is sufficient for the estimation of the gradient.
Although Xue's learning rule can find the optimal strategy by using only the information of the parent's type, the relationship between Xue's rule and the gradient is unclear.
The information of the parent's type might be insufficient to estimate gradient, and communications between agents at the same generation might be required.
The sufficient condition is also important for engineering systems to find new variants of the genetic algorithm and the particle filter.

The last one is how to quantify and predict the acceleration of natural selection by learning.
For the conventional evolutionary processes with natural selection, we have \emph{Fisher's fundamental theorem (FF-thm)} and its variants~\cite{fisher1930genetical}.
The theorem states that the increase in the mean fitness of a population is proportional to the variance of the fitness in the population. 
From the relation, we can predict the progress and speed of the evolution in the population.
Because the evolutionary process becomes more complicated by taking learning of individual agents into account, a simple relationship similar to FF-thm would facilitate our understanding of the impact and efficiency of learning.
Furthermore, such a relationship may be applied to analyzing the performance of engineering systems.

In this paper, we  address the three problems.
First, we propose \emph{ancestral learning}, which utilizes only the information  transmitted from the ancestors via epigenetic states or memes.
The ancestral learning is simple and therefore biologically reasonable, which also generalizes Xue's learning rule.
We validate that the ancestral learning accelerates the evolutionary process by numerically showing that the optimal type-switching strategy is acquired by the ancestral learning faster than by the zero-th order mutational rules.
Second, we prove that the ancestral information is sufficient to estimate the gradient of fitness.
In particular, we show that ancestral learning updates the strategy into the direction of the gradient.
Third, we derive an extended FF-thm for the ancestral learning, which relates the variation of fitness among ancestors to the fitness gain by the ancestral learning.
With this theorem, we can predict the acceleration of evolutionary processes by ancestral learning, which depends on the property of environment. The theorem enables us to  quantitatively understand when and why ancestral learning becomes beneficial.

\section{Setup}
\label{sec:setup}

\begin{figure}[t]
  \begin{center}
        \includegraphics[width= \linewidth]{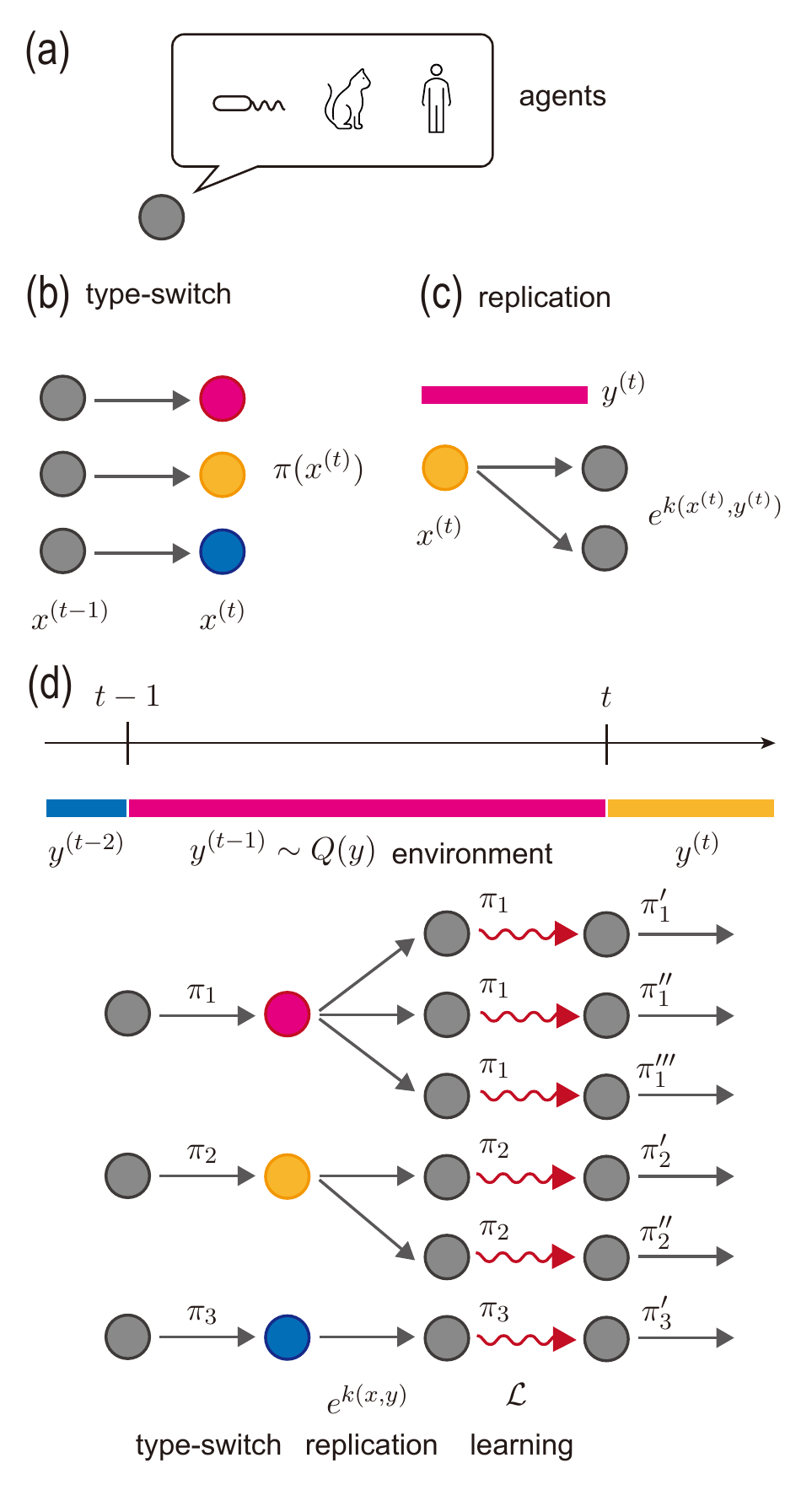}
   \end{center}
    \caption{Schematic representation of the setup for learning in evolutionary processes. 
    We will consider agents that can replicate and learn. Examples are microbes, animals, and humans (a). 
    (b--d) Schematic illustrations of the model. An agent at time $t-1$ first determines its type based on its strategy $\pi$ (b). 
    The agent then produces $e^{k(x,\suptime{y}{t-1})}$ daughters depending on its type $x$ and the environmental state $\suptime{y}{t-1}$ (c). 
    After the replication, the daughters inherit strategies updated by a given learning rule $\learningRule$ (d).}
    \label{fig:model}
\end{figure}

We consider population dynamics of asexual agents with a discrete generation time $t \in \{0,1,2,\dots\}$ (Fig.~\ref{fig:model} (b)).
Let $\suptime{x}{t} \in \Xset$ and $\suptime{y}{t} \in \Yset$ be the type of an agent and the state of the environment at time $t$.
The type models the phenotypic trait of organisms in biological systems.
Each agent has its own stochastic type-switching strategy  $\pif \in \R^\Xset$ where $\pif(x)$ is the probability to switch into type $x$ and $\pif$ satisfies that $\sum_{x \in \Xset} \pif(x) = 1$ and $\pif(x) \ge 0$ for all $x \in \Xset$ (Fig.~\ref{fig:model} (b)).
We call $\pif$ a strategy and $\suptime{y}{t} \in \Yset$ the environmental state at time $t$.
Environmental state $\suptime{y}{t}$ at time $t$ follows a distribution $\Q(y)$ on $\Yset$, which is independent of $t$.
An agent with type $x$ under environmental state $y$ duplicates asexually and produces $e^{k(x,y)}$ daughters on average (Fig.~\ref{fig:model} (c)).
The term $e^{k(x,y)}$ is called an \emph{individual fitness}~\footnote{Precisely, the individual fitness is the mean number of the daughters. We use the term ``fitness'' to follow the terminology of the FF-thm~\cite{fisher1930genetical}.} of the agent.
We define paths (histories) of the types along a lineage and the environmental states from time $0$ to time $t-1$ as $\suptime{\Xpath}{t}$ and $\suptime{\Ypath}{t}$, respectively.

To define a ``fitness'' of strategy, we first consider the case where the agents cannot learn the strategy and the strategy $\pif$ is fixed in a population and over generations.
The number $\suptime{N}{t}$ of the agents in the population at time $t$ under a path $\suptime{\Ypath}{T}$ of environmental states  becomes
\begin{align}
    \label{eq:model_PD}
    \suptime{N}{t}_{\pif}[\suptime{\Ypath}{T}] = \left[\sum_{x \in \Xset} e^{k(x, \suptime{y}{t-1})} \pif(x) \right]\suptime{N}{t-1}_{\pif}[\suptime{\Ypath}{T}].
\end{align}
Here the initial size $\suptime{N}{0}$ of the population is given as an initial condition.
When the dependency on strategy $\pif$ or  a path  $\suptime{\Ypath}{T}$ of environment states is clear from the context, we omit them.
We can use this dynamical system to define a ``fitness'' of the strategy $\pif$.
The cumulative population fitness of  strategy $\pif$ under  $\suptime{\Ypath}{t}$ up to  time $t$ is defined as
\begin{align}
    \label{eq:def_grwoth_rate}
    \suptime{\Lambda}{t}(\pif \mid \suptime{\Ypath}{t}) = \log \frac{\suptime{N}{t}_{\pif}[\suptime{\Ypath}{t}]}{\suptime{N}{0}_{\pif}}.
\end{align}
The time-averaged population fitness of $\pif$ is defined as
\begin{align}
    \label{eq:def_fitness}
    \lambda(\pif) = \lim_{t \to \infty} \frac{1}{t} \suptime{\Lambda}{t}(\pif \mid \suptime{\Ypath}{t}),
\end{align}
which exists almost surely and independently of $\suptime{\Ypath}{t}$ owing to the ergodicity of the environmental state~\cite{seppalainen1994large, kifer1996perron}.
In the following, we call $\lambda(\pif)$ the population fitness in short~\footnote{The population fitness is also called a growth rate.}.

When the agents learn their strategy (Fig.~\ref{fig:model} (d)), the number $\suptime{N}{t}(\pi)$ of the agents with a strategy $\pi$ at time $t$ becomes
\begin{align}
    \label{eq:model-PD-learn}
    \suptime{N}{t}(\pi) = \left[\sum_{x \in \Xset, \pi'} \learningRule(\pi \mid \pi')  e^{k(x, \suptime{y}{t-1})} \pi'(x) \right]\suptime{N}{t-1}(\pi').
\end{align}
Here, $\learningRule(\pi \mid \pi')$ is a (possibly stochastic) learning rule, which satisfies $\sum_{\pi} \learningRule(\pi \mid \pi') = 1$.
The learning rule $\learningRule$ can depend on available information for agents to learn.
We consider the following sources of the available information.
Each agent can transmit information to the next generation via epigenetic states or memes.
Specifically, each agent can access the frequency of the types that ancestors chose.
While we do not explicitly consider communications between agents, we show that learning rule without communication is sufficient to achieve acceleration of evolutionary processes via estimating fitness gradient.
In addition,  we do not assume that the agent can sense the environmental state $y$.
For a further generalization on these assumptions, see Section~\ref{sec:discussion}.

Under this setting, we consider how agents in the population can gradually acquire the optimal strategy:
\begin{align}
    \pi^{*} = \argmax_{\pi} \lambda(\pi),
\end{align}
by the zeroth or the first order learning.
We note that the optimal strategy is unique due to the concavity of the population fitness $\lambda(\pi)$ to $\pi$.
The concavity of $\lambda$ follows from Eq.~\eqnref{eq:lambda-explicit} that we prove later.

\section{Ancestral Learning}
\label{sec:ancestral-learning}

We first propose \emph{ancestral learning} and validate that it can accelerate the evolutionary process.
 Ancestral learning is self-reinforcement of strategy by positive feedback.
It updates strategy every $\timeEstimateRetro$ generations, where $\timeEstimateRetro$ is a hyperparameter called an \emph{update interval}.
We suppose that the update occurs at time $t = i \timeEstimateRetro - 1$ ($i = 1,2,\dots$).
We specially regard that the initial strategies $\suptime{\pif}{0}$ are acquired at time $-1$ by the zero-th update.
After an agent at time $(i-1)\timeEstimateRetro - 1$ acquires the strategy $\suptime{\pif}{i-1}$ by the $(i-1)$-th update,  its descendants at time $t'$ $((i-1)\timeEstimateRetro \le t' < i \timeEstimateRetro)$ have the same strategy.
At time $i\timeEstimateRetro - 1$, i.e., at the next update,
each of the descendants calculates the empirical distribution $\retroEmp^{\suptime{\pif}{i-1}}$ of the ancestor's types back to time $(i-1)\timeEstimateRetro$.
In particular, $\retroEmp^{\suptime{\pif}{i}}(x) := \sum_{t' = (i-1) \timeEstimateRetro }^{i \timeEstimateRetro - 1} \delta_{x, \suptime{x}{t'}}$, where $\delta_{x,x'}$ is the Kronecker's delta and $\suptime{x}{t'}$ is the type of the ancestor at time $t'$.
When $\suptime{\pif}{i-1}$ is clear from the context, we omit it. 
After obtaining the empirical distribution, the agent updates its strategy by a rule 
\begin{align}
    \label{eq:update-anecstral-learning}
\suptime{\pif}{i} \gets (1 - \learningRate) \suptime{\pif}{i-1} + \learningRate \retroEmp^{\suptime{\pif}{i-1}}
\end{align}
where $\learningRate$ is a hyperparameter called a \emph{learning rate}.
In this rule of ancestral learning, the strategy after $i-1$th update $\suptime{\pif}{i}$ is the a mixture of the previous strategy $\suptime{\pif}{i-1}$ with the frequency of types that ancestors chose $\retroEmp^{\suptime{\pif}{i-1}}$. If the learning rate is close to $1$, i.e., $\learningRate\approx 1$, the updated strategy $\suptime{\pif}{i}$ becomes identical to the ancestor's type frequency. If $\learningRate\approx$ is small, the information of ancestor's type is gradually assimilated to the strategy.
Ancestral learning coincides with Xue's rule~\cite{xue2016evolutionary} when $\timeEstimateRetro = 1$.
In addition, the rule does not require communications between agents.

\begin{figure}[t]
  \begin{center}
        \includegraphics[width= \linewidth]{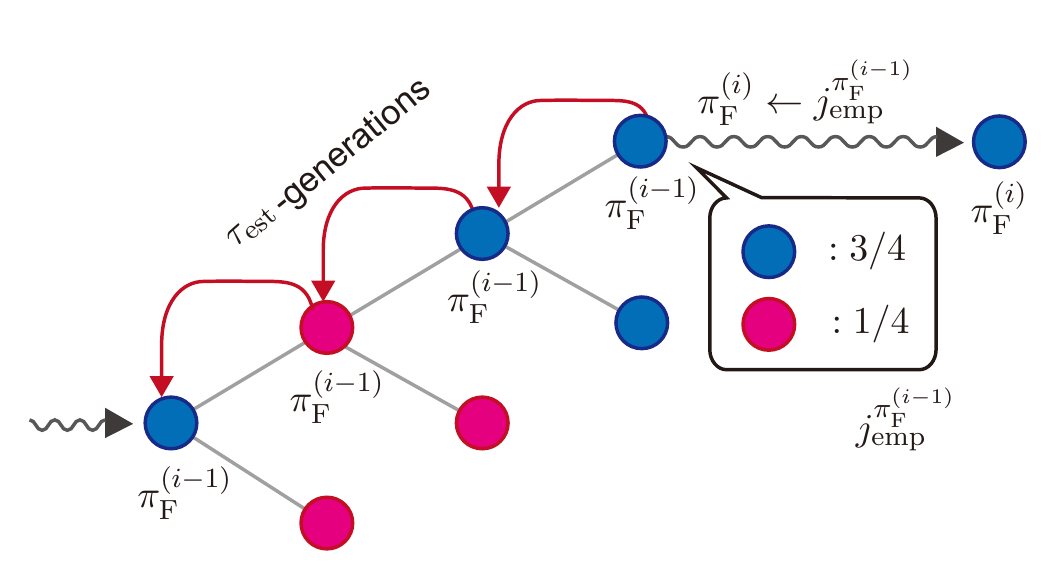}
   \end{center}
    \caption{Schematic representation of ancestral learning. After an agent and acquires the strategy $\suptime{\pif}{i-1}$ by the $(i-1)$-th update, its descendants have the same strategy for $\timeEstimateRetro$-generations.
    At the next update, each of the descendants calculates the empirical distribution $\retroEmp^{\suptime{\pif}{i-1}}$ and updates the strategy  by Eq.~\ref{eq:update-anecstral-learning}. The figure corresponds to the case where $\alpha = 1.0$ and $\timeEstimateRetro = 4$.}
    \label{fig:ancestral_learning}
\end{figure}

Ancestral learning is a biologically reasonable learning rule.
The information used in the rule is only the empirical distribution $\retroEmp$ of the ancestor's types, which can be stored and transmitted via epigenetic states or memes as we discussed before.
Owing to this property, we call $\retroEmp$ ancestral information.
Also, the memory to store $\retroEmp$ is reasonably small.
In the following, we prove that the compressed information $\retroEmp$ instead of the whole path $\suptime{\Xpath}{t}$ of the acestor's type is sufficient for attaining the optimal strategy. 
The update rule of ancestral learning seems natural since it is similar to Hebb's rule~\cite{hertz1991introduction} as pointed out in~\cite{xue2016evolutionary}.
Hebb's rule is a self-reinforcement by positive feedback in that the synaptic connection between activated and coactivated neurons are strengthened.

The intuitive explanation why ancestral learning can attain the optimal strategy is that replicating the types of the survived ancestors is likely to contribute to the survival of the descendants.
Due to the growth competition among the population,
the empirical distribution $\retroEmp$ of ancestor's types deviates from the strategy $\pif$,
and $\retroEmp$ seen as a strategy has a greater population fitness than $\pif$.
This deviation known as \emph{survivorship bias} works as the driving force of ancestral learning.

To see the intuition more precisely, let us consider the simple case where the environment is constant $\Yset = \{*\}$, the learning rate $\alpha = 1.0$, and the update interval $\timeEstimateRetro$ is sufficiently long.
In this case, the individual fitness only depends on type and we can omit $y$ in $e^{k(x,y)}$ as $e^{k(x)}$.
The optimal strategy $\pif^*$ is 
\begin{align}
    \begin{cases}
        \pif^*(x^*) = 1 & (x^* = \argmax_{x \in \Xset} k(x)),\\
        \pif^*(x) = 0 & (\text{otherwise}),
    \end{cases}
    \label{eq:optimal-strategy-const-env}
\end{align}
which means that $\pif^*$ always selects the type $x^{*}$ maximizing the individual fitness $e^{k(x)}$.
We calculate $\retroEmp^{\suptime{\pif}{i-1}}$ to see how ancestral learning updates the strategy $\suptime{\pif}{i-1}$ and to check that $\suptime{\pif}{i}$ converges to the optimal $\pif^{*}$ as $i \to \infty$.
Since $\retroEmp(x) =  \sum_{t' = (i-1) \timeEstimateRetro}^{i \timeEstimateRetro - 1} \delta_{x, \suptime{x}{t'}}$ is the sum of independent and identically distributed random variables $\{\delta_{x, \suptime{x}{t'}}\}_{t'}$, the law of large numbers implies that 
\begin{align}
    \label{eq:expected-retroEmp}
    \retroEmp \approx \average{\delta_{x, \suptime{x}{t'}}}{},
\end{align}
when $\timeEstimateRetro$ is sufficiently long (we discuss the case when $\timeEstimateRetro$ is not large, and show that small learning rate $\learningRate$ can compensate small $\timeEstimateRetro$ in Sec.~\ref{sec:tradeoff}).
We can interpret $\average{\delta_{x, \suptime{x}{t'}}}{}$ as the following probability.
Recall that an agent at time $(i-1)\timeEstimateRetro -1$ acquires $\suptime{\pif}{i-1}$ via the update by ancestral learning and that its descendants have the same strategy until the next update at time $i \timeEstimateRetro - 1$.
Let us consider the sub-population that consists of the descendants.
We choose an agent at time $t'+1$ ($(i-1)\timeEstimateRetro \le t' < i \timeEstimateRetro$) from the sub-population uniformly at random.
Under this setting, $\average{\delta_{x, \suptime{x}{t'}}}{}$ is the probability $\pib(x)$ that the parent of the chosen agent expresses type $x$.
Let $\suptime{N'}{t'}$ be the number of the agents in the sub-population at time $t'$.
In the sub-population, the number of the agents at time $t'+1$ whose parent expresses type $x$ is $e^{k(x)} \suptime{\pif}{i-1}(x) \suptime{N'}{t'}$ and the total number of the agents at time $t'+1$ is $\sum_{x \in \Xset}e^{k(x)}\suptime{\pif}{i-1}(x)\suptime{N}{t'}$.
Therefore, the probability $\pib(x)$ is
\begin{align}
    \pib(x) &= \frac{e^{k(x)}\suptime{\pif}{i-1}(x)\suptime{N'}{t'}}{\sum_{x \in \Xset} e^{k(x)}\suptime{\pif}{i-1}(x)\suptime{N'}{t'}} \nonumber\\
    &= \frac{e^{k(x)}\suptime{\pif}{i-1}(x)}{\sum_{x \in \Xset} e^{k(x)}\suptime{\pif}{i-1}(x)}.
    \label{eq:retro-const-env}
\end{align}
This equation and Eq.~\eqnref{eq:expected-retroEmp} imply that  $\retroEmp \approx \pib$ when $\timeEstimateRetro$ is sufficiently large.
The probability $\pib$ is called a \emph{retrospective process} of $\suptime{\pif}{i-1}$ for the constant environment~\cite{hermisson2002mutation, baake2007mutation, georgii2003supercritical, sughiyama2015pathwise}.
The retrospective process is biased so that $\pib(x^*)$, the probability to switch into the optimal type $x^*$, is larger than $\suptime{\pif}{i-1}(x^*)$ and therefore better fitted to the environment.
Since ancestral learning updates the strategy to $\retroEmp$, the strategy becomes $\suptime{\pif}{i}(x) \propto e^{ik(x)} \suptime{\pif}{0}(x)$ after the $i$-th update.
Consequently, $\suptime{\pif}{i} \to \pi^*$ as $i \to \infty$.

We next consider the case where the environment is not constant.
We calculate $\retroEmp^{\suptime{\pif}{i}}$ as with the constant-environment case.
Since the environmental state $\suptime{y}{t}$ follows $Q(y)$ independently of $t$, 
the law of large numbers implies that
\begin{align}
    \label{eq:expected-retroEmp-random-env}
    \retroEmp(x) \approx \average{\average{\delta_{x, \suptime{x}{t'}} \mid y}{} }{Q(y)},
\end{align} where $\average{\delta_{x, \suptime{x}{t'}} \mid y}{}$ is the conditional expectation of  $\delta_{x, \suptime{x}{t'}}$ given that the environmental state at time $t'$ is $y$.
We can interpret $\average{\delta_{x, \suptime{x}{t'}} \mid y}{}$ as the following conditional probability $\pif(x \mid y)$.
Let us consider an agent that acquires strategy $\suptime{\pif}{i-1}$ at time $(i-1)\timeEstimateRetro - 1$ and the sub-population that consists of its descendants as before.
Suppose that we choose an agent at time $t'+ 1$ $((i-1)\timeEstimateRetro \le t' < i\timeEstimateRetro)$ from the population  and that the environmental state at time $t'$ is $y$.
Under this setting, $\average{\delta_{x, \suptime{x}{t'}} \mid y}{}$ is the probability $\pib(x \mid y)$ that the parent of the chosen agent expresses type $x$.
By a similar argument, we have
\begin{align}
    \pib(x \mid y) = \frac{e^{k(x,y)} \pif(x)}{\sum_{x' \in \Xset} e^{k(x',y)} \pif(x') }.
\end{align}
The probability $\pib(x \mid y)$ is also called the retrospective process and fitted to the environmental state $y$ better than $\pif$.
This equation and Eq.~\eqnref{eq:expected-retroEmp-random-env} imply that $\retroEmp(x)$ converges to  the averaged retrospective process $\pibm(x): = \sum_{y} \pib(x \mid y) Q(y)$.
Therefore, $\pif$ is updated to the mixture of the strategies $\pib(x \mid y)$, each of which is better fitted to the corresponding environmental state.
We will numerically (Section~\ref{sec:simulation-ancestral-learning}) and theoretically (Section~\ref{sec:universal}) prove that update to such a mixture strategy leads to the optimal one.

\section{Ancestral learning can accelerate the evolutionary processes}
\label{sec:simulation-ancestral-learning}

Next, we validate that learning can accelerate the evolutionary process by numerically showing that the optimal type-switching strategy is acquired with ancestral learning faster than with the zero-th order mutational rule.

We simulate the evolutionary process by a multi-type branching process in a random environment~\cite{harris1964theory, weissner1971multitype}.
Namely, we simulate the dynamical system defined by Eq.~\eqnref{eq:model-PD-learn} while taking the individuality and finite number of the agents into account.
In the simulation, we set $\Xset = \Yset = \{0,1,2\}$ (Fig.~\ref{fig:ancestral_learning} (a)).
In the following, the colors are numbered from left to right in the figure.
Namely, $0 \in \Yset$ corresponds to red, $1$ to yellow, and $2$ to blue as in Fig.~\ref{fig:ancestral_learning} (a).
We also set $Q(0) = 0.6$ and $Q(y) = 0.2$ for the other $y \in \Yset$.
Each agent with type $x$ under environmental state $y$ has four daughters if $x = y$ and one daughter, otherwise.
In short, $e^{k(x,y)} = 4$ if $x = y$ and $e^{k(x,y)} = 1$, otherwise.
We represent strategy $\pif$ as a vector of the form $(\pif(0), \pif(1), \pif(2))$.
Due to the symmetry of $e^{k(x,y)}$,  the zero-th component  $\pi^*(0)$ of the optimal strategy is higher than the others.
We start the simulation from a single agent, whose
 initial strategy is $\suptime{\pif}{0} = (1/3, 1/3, 1/3)$.
We limit the number of the agents in the population to $\Nmax = 30$ to avoid the intractability of the numerical experiment due to the exponential growth of the number of the agents.
If the number of the agents in the next generation exceeds $\Nmax$, we select $\Nmax$ agents uniformly at random.

We investigate three learning rules.
Each learning rule updates the strategy at every time step, i.e., $\timeEstimateRetro = 1$.
The first learning rule is ancestral learning with learning rate $\learningRate = 0.01$.
The second and the third ones are the zero-th order mutational rules.
Since there are innumerable zero-th order learning rules, we choose two representative ones
to perform the control experiments for ancestral learning.
The second learning rule is
$\pif' \gets (1 - \alpha) \pif + \learningRate \delta_{x,x_k}$,
where $\pif$ and $\pif'$ are the strategies of the agent before and after the update, respectively, and $x_k$ is chosen uniformly at random from $\Xset$.
In biological systems, this rule can be seen as a random  mutation of $\pif$ whose rate is constant. 
The trajectory of $\pif$ updated by this rule is a random walk over $\R^\Xset$ if no growth occurs, that is, $e^{k(x,y)} = 1$ for all $x \in \Xset$ and $y \in \Yset$.
We therefore call this learning rule a random walk.
The third learning rule is
$\pif' \gets (1 - \alpha) \pif + \learningRate \delta_{x,x_k}$, 
where $x_k$ is sampled from the discrete distribution $\pif$.
In biological systems, this rule can be seen as the mutation of $\pif$ whose rate is dependent on the current $\pif$.
The change of mutation rate is known as an adaptive mutation~\cite{rosenberg2001evolving}.
Therefore, we call this learning rule an adaptive random walk.
The adaptive random walk coincides with ancestral learning if no growth occurs.
In this sense, the adaptive random walk is a control to see the effect of the population growth on ancestral learning.

\begin{figure*}[t]
  \begin{center}
        \includegraphics[width=0.8 \linewidth]{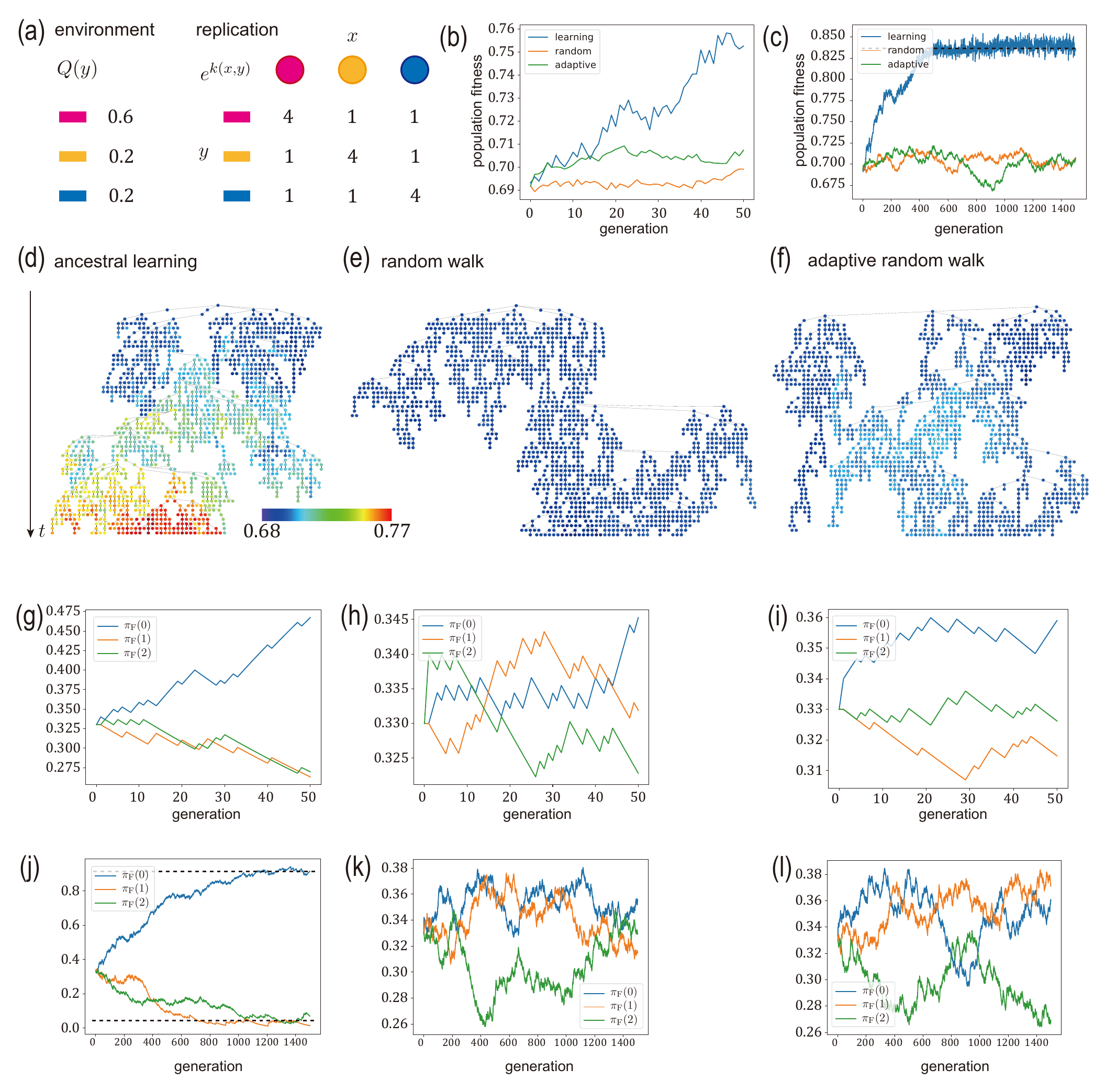}
   \end{center}
    \caption{Numerical experiments of ancestral learning. (a) The parameters of the model. In the panel, $0 \in \Yset$ corresponds to red, $1$ to yellow, and $2$ to blue. The red environmental state occurs more frequently than the others. An agent has more daughters if its type is equal to the environmental state. (b--l)  The simulated lineage trees of the agents that adopt ancestral learning (d), the random walk (e), and the adaptive random walk (f). Each curve in (b) shows the trajectory of the population fitness $\lambda$ along the lineage of the most successful agent, whose $\lambda$ is the maximum of the agents at the end of each lineage tree. Ancestral learning increases the population fitness the best among the three learning rules. The curve (c) shows the same plot as (b) for another simulation until $t = 1500$. The dotted line shows the population fitness of the most successful agent with ancestral learning at $t = 1500$. By the longer simulation, we can see the convergence of the population fitness in the population with ancestral learning.
    In (d--f), each point corresponds to an agent and its color represents the population fitness $\lambda$ of the agent. [
    Black lines connect parents to their daughters. 
    The curves (g-i) show the trajectories of the strategy $\pif$ along the lineage of the most successful agent updated by ancestral learning (f), by the random walk (g), and by the adaptive random walk (h), respectively.
    The curves (g--i) are truncations of those in (j--l) up to $t = 50$. 
    In (j), the upper dotted line shows $\pif(0)$ of the most successful agent and the lower one is the average of $\pif(1)$ and $\pif(2)$.
    We can see that the strategy of the most successful agent with ancestral learning converges to $(0.92, 0.04, 0.04)$ approximately. The converged strategy is close to the optimal since it satisfies the optimality condition (\cite[Theroem 16.2.1]{cover1999elements}
    ) with small error.
    }
    \label{fig:lineage-tree-learning}
\end{figure*}

Figure~\ref{fig:lineage-tree-learning} is the result of the simulation of the three learning rules, which shows that ancestral learning accelerates the evolutionary process.
We show lineage trees up to $t = 50$.
The population fitness of the population with ancestral learning increases faster than those with the other learning rules (Fig.~\ref{fig:ancestral_learning} (b)) along the lineage of the most successful agent, whose population fitness is the maximum of the agents at the end of each lineage tree.
The acceleration of the evolutionary process is also observed at the lineage tree level  (Fig.~\ref{fig:ancestral_learning} (d--f)).
In Fig.~\ref{fig:ancestral_learning} (g--i), we select the lineage of the most successful agent in each lineage tree and plot the trajectory of $\pif$ along the lineage.

To see whether the optimal strategy is acquired by ancestral learning, we run another simulation until $t = 1500$.
We first checked that the strategy converges, i.e., the strategy before and after the update is almost identical when $t$ is sufficiently large.
We then verified that the converged strategy is the optimal strategy.
We checked the convergence of the strategy along the lineage of the most successful agent with ancestral learning.
The strategy converges since the population fitness along the lineage reaches a ceiling (Fig.~\ref{fig:ancestral_learning} (c)).
The convergence is also supported from the trajectory of the strategy ( Fig.~\ref{fig:ancestral_learning} (j--l)).
The converged strategy (approximately $(0.92, 0.04, 0.04)$) of the most successful agent with ancestral learning is close to the optimal since it satisfies the optimality condition 
(Karush-Kuhn-Tucker condition) with small error~\cite[Theorem 16.2.1]{cover1999elements}.

From these results, we conclude that ancestral learning accelerates the evolutionary process.
Since ancestral learning does not use the information via a communication between the agent at the same time, we numerically showed that learning can accelerate the evolutionary process even without communications.

\section{Ancestral Information Is Sufficient to Estimate Gradient}
\label{sec:universal}

We next address the second problem: whether an agent can estimate the gradient of the population fitness or not.
Although we numerically showed that ancestral learning accelerates the evolutionary process, the relationship between ancestral learning and the fitness gradient is unclear.
The ancestral information $\retroEmp$ used in ancestral learning might be insufficient to estimate the gradient and the communication between the agents at the same generation might be required.
In this section, we prove that the ancestral information $\retroEmp$  is sufficient to estimate the gradient.
It theoretically implies that an agent can estimate the gradient without the communication between agents. 
It also implies that ancestral learning updates the strategy into the direction of the gradient.

To calculate the gradient of the population fitness, we employ a \emph{pathwise formulation} and \emph{variational principle}~\cite{sughiyama2015pathwise} of the population dynamics.
Let us consider the case where the path of the environmental state is $\suptime{\Ypath}{t}$ and the agents do not learn and stick to a fixed strategy $\pif$.
By applying Eq.~\eqref{eq:model_PD} recursively, we know that the number $N_{\pif}[\suptime{\Xpath}{t} \mid \suptime{\Ypath}{t}]$ of the agents at time $t$ whose path of the type of the ancestors is $\suptime{\Xpath}{t}$ satisfies
\begin{align}
    N_{\pif}[\suptime{\Xpath}{t} \mid \suptime{\Ypath}{t}] = e^{k(\suptime{\Xpath}{t}, \suptime{\Ypath}{t})} \forward[\suptime{\Xpath}{t}] \suptime{N}{0}(\suptime{x}{0}),
\end{align}
where $\suptime{N}{0}(x)$ is the number of the initial agent with type $x$ and the quantities $k(\suptime{\Xpath}{t}, \suptime{\Ypath}{t}) := \sum_{t'=0}^{t-1} k(\suptime{x}{t'}, \suptime{y}{t'})$ and $\forward[\suptime{\Xpath}{t}] = \prod_{t'=0}^{t-1} \pif(\suptime{x}{t'})$ are the pathwise (historical) individual fitness and pathwise forward probability, respectively.
Under the pathwise forumulation, we can represent the cummulative population fitness as
\begin{align}
    \suptime{\Lambda}{t}(\pif \mid \suptime{\Ypath}{t}) = \log \sum_{\suptime{\Xpath}{t}} e^{k(\suptime{\Xpath}{t}, \suptime{\Ypath}{t})} \forward[\suptime{\Xpath}{t}].
\end{align}
Since each $\suptime{y}{t}$ follows $Q(y)$ independently, the population fitness satisfies (cf.~\cite{seppalainen1994large, kifer1996perron})
\begin{align}
    \label{eq:lambda-explicit}
    \lambda(\pif) = \average{\log \average{e^{k(x, y)}}{\pif(x)}}{Q(y)}.
\end{align}
The form of $\log \average{\cdot}{\pif}$ in the right hand side is equivalent to the scaled cummulant generating function~\cite{dembo2010large} and the following variational principle holds:
\begin{align}
    \label{eq:variation-growth-rate}
    \lambda(\pif) = \average{ \max_{\pi}\left\{\sum_{x \in \Xset} k(x)\pi(x) - \KL{\pi}{\pif} \right\} }{Q(y)},
\end{align}
where $\pi$ runs over all distributions on $\Xset$ and $\KL{\cdot}{\cdot}$ is the \emph{ Kallbuck-Leibler divergence (KL-divergence)} defined by
\begin{align}
     \KL{\pi}{\pi'} := \sum_{x \in \Xset} \pi(x) \log \frac{\pi(x)}{\pi'(x)}.
\end{align}
See Appendix~\ref{subsec:variational-rep-derivation} for the proof.
By direct calculation, we can see that the maximizer is $\pibm$.
We can calculate the derivative of the population fitness from the variational principle:
\begin{align}
    \label{eq:derivative-lambda}
    \frac{\partial \lambda(\pif(x))}{\partial \pif(x)} = \frac{\pibm(x)}{\pif(x)}.
\end{align}
See Appendix~\ref{subsec:gradient-growth-rate} for the proof.

We now have all the ingredients to calculate the gradient of the population fitness.
Since the strategy $\pif$ has a constraint $\sum_{x \in \Xset} \pif(x) = 1$, we consider the following definition of the gradient.
A \emph{gradient} at $\pif$ under the constraint  $\sum_{x \in \Xset} \pif(x) = 1$ is defined by
\begin{align}
    \lim_{\epsilon \to^+ 0} \argmax_{\substack{\delta \pi \\ \pif + \delta \pi \in \sphere{\pif}{\epsilon}}} \left\{\lambda(\pif + \delta \pi)\right\},
\end{align}
where the limit is one-sided from the positive real numbers, $\delta \pi \in \R^{\Xset}$ with $\sum_{x \in \Xset} ( \pif(x) + \delta \pi(x) ) =  1$, i.e., $\sum_{x \in \Xset} \delta \pi(x) = 0$, and $\sphere{\epsilon}{\pif}$ is the sphere around $\pif$ with radius $\epsilon$.
To define the sphere, we use the KL-divergence as a natural distance over distributions on $\Xset$.
Intuitively, the gradient is the direction into which the population fitness increases the most among all alternatives that satisfy the constraint and have the same infinitesimal distance from $\pif$.
The definition is related to a \emph{proximal operator}~\cite{parikh2014proximal} and coincides with the usual gradient if no constraint is imposed and the sphere is defined by the Euclidean distance.
We prove that the gradient is directed toward $\pibm$, i.e.,
\begin{align}
    \label{eq:gradient-pib}
    \lim_{\epsilon \to^+ 0} \argmax_{\substack{\delta \pi \\ \pif + \delta \pi \in \sphere{\pif}{\epsilon} }} \left\{\lambda(\pif + \delta \pi)\right\} \propto \pibm - \pif.
\end{align}
See Appendix~\ref{subsec:gradient-growth-rate} for the proof.

The result addresses the second problem.
To estimate the gradient, an agent must estimate $\pibm$.
By the discussion in the last paragraph of Section~\ref{sec:ancestral-learning},  the ancestral information $\retroEmp$ is the unbiased estimator of $\pibm$, that is, $\average{\retroEmp}{} = \pibm$.
Therefore, an agent can estimate the gradient from ancestral information without communication between the agents at the same generation.
The explicit formula of the gradient also implies that ancestral learning updates the strategy into the direction of the gradient.
The direction $\suptime{\pif}{i+1} - \suptime{\pif}{i}$ of the update of the strategy by ancestral learning equals the right hand side of Eq.~\eqnref{eq:gradient-pib} on average.
In particular, ancestral learning finds the optimal strategy if the learning rate is sufficiently small since $\lambda$ is concave.

\section{Fisher's Fundamental Theorem for Ancestral Learning}
\label{sec:ff-thm}
We address the last problem, the quantification of the acceleration of the evolutionary process by  learning, via extending the FF-thm to ancestral learning.
Ancestral learning may increase the population fitness much faster under some environments than others depending the stochastic property $Q(y)$ of the environments.
In addition, the acceleration might also depend on the update interval $\timeEstimateRetro$ and the learning rate $\learningRate$.
We can understand such dependency as well as when and why learning becomes beneficial by extending the conventional FF-thm to ancestral learning.

Let us first review the conventional FF-thm for natural selection~\cite{fisher1930genetical}.
The FF-thm relates the speed of the evolution and the variance of the individual fitness in the population.
To illustrate this, we consider the following fixed-type population dynamics in a constant environment.
The set of types is $\Xset$ as before.
The type of the daughter is the same as that of the parent.
The environment is constant $\Yset = \{*\}$.
The individual fitness of type $x$ is $e^{k(x)}$.
Here, we omit the dependency  of the individual fitness $e^{k(x,*)}$ on the environmental state $*$ since the environment is constant.
Under this setting, the number $\suptime{N}{t}(x)$ of the agent with type $x$ at time $t$ is 
\begin{align}
    \label{eq:model-fisher-PD}
    \suptime{N}{t}(x) = e^{k(x)}\suptime{N}{t-1}(x).
\end{align}
Since we are interested in statistics of the population such as the variance of the individual fitness, we focus on the fraction $\suptime{p}{t}(x)  := \suptime{N}{t}(x) /\sum_{x \in \Xset} \suptime{N}{t}(x)$ of the agent with type $x$ at time $t$ instead of $\suptime{N}{t}(x)$.
The time evolution of $\suptime{p}{t}$ derived from Eq.~\eqref{eq:model-fisher-PD} is
\begin{align}
    \label{eq:model-fisher-PD-fraction}
    \suptime{p}{t}(x) = \frac{e^{k(x)}\suptime{p}{t-1}(x)}{ \sum_{x' \in \Xset} e^{k(x)}\suptime{p}{t-1}(x)}.
\end{align}
We define a \emph{covariance} of random variables $f(x)$ and $g(x)$ with respect to a probability distribution $p(x)$ over $\Xset$ by
\begin{align}
    \cov{f(x)}{g(x)}{p} := \average{f(x)g(x)}{p} - \average{f(x)}{p}\average{g(x)}{p}.
\end{align}
From this, a \emph{variance} is also defined as 
\begin{align}
    \V_{p}[f(x)] := \cov{f(x)}{f(x)}{p}.
\end{align}
One of the measures of the evolutionary speed is the gain of the the mean individual fitness $\average{e^{k(x)}}{\suptime{p}{t}}$.
The gain satisfies the following relation due to Eq.~\eqnref{eq:model-fisher-PD-fraction}:
\begin{align}
    \label{eq:FF-thm-mean-fitness}
    \Delta \average{e^{k(x)}}{\suptime{p}{t}} &:= \average{e^{k(x)}}{\suptime{p}{t}} - \average{e^{k(x)}}{\suptime{p}{t-1}}\\
    &= \V_{\suptime{p}{t-1}}\left[e^{k(x)} \right] / \average{e^{k(x)}}{\suptime{p}{t-1}}.
    \label{eq:ff-thm-original}
\end{align}
See Appendix~\ref{subsec:ff-thm-derivation} for the proof.
The equation reveals the relationship between the evolutionary speed and the variance of the individual fitness in the population.
This equation is called the FF-thm for natural selection~\footnote{The FF-thm for natural selection is sometimes formulated in a continuous time model. We here present a discrete time version to see the connection to our ancestral learning. }.

Since we are not interested in the mean individual fitness but the population fitness 
\begin{align}
    \suptime{\lambda}{t} := \log ({\sum_{x \in \Xset} \suptime{N}{t}(x)})/({\sum_{x \in \Xset} \suptime{N}{t-1}(x)})
\end{align}
at time $t$, we present an FF-thm of the population fitness.
We introduce variants of the covariance and variance to extend the conventional FF-thm.
We define a \emph{log-covariance} and a \emph{log-variance} by
\begin{align}
    \logCov{f(x)}{g(x)}{p} := \log \frac{\average{f(x)g(x)}{p}}{\average{f(x)}{p}\average{g(x)}{p}}, \\
    \logV_p[f(x)] := \logCov{f(x)}{f(x)}{p},
\end{align}
respectively.
The log-covariance measures the similarity of two random variables as the covariance does since the log-covariance is monotonically increasing with respect to the covariance.
Indeed, we can prove that
\begin{align}
    \label{eq:covariance-and-log-covariance}
    \logCov{f(x)}{g(x)}{p} = \log\left(         \frac{\cov{f(x)}{g(x)}{p} }
            {\average{f(x)}{p} \average{g(x)}{p} }
        + 1 \right),
\end{align}
by direct calculation.
By using these quantities, we can obtain an extended FF-thm for the population fitness by a similar argument to Eq.~\eqnref{eq:ff-thm-original}:
\begin{align}
    \label{eq:FF-thm-growth-rate}
    \Delta \suptime{\lambda}{t} := \suptime{\lambda}{t} - \suptime{\lambda}{t-1} = \logV_{\suptime{p}{t-1}}\left[e^{k(x)}\right]
\end{align}
See Appendix~\ref{subsec:ff-thm-derivation} for the proof.
This equation reveals the relationship between the speed of the evolutionary process measured by the gain of the population fitness and the log-variance of the individual fitness in the population.

The FF-thm for the population fitness has a close connection to ancestral learning.
To see this, let us first consider a simple case where the environment is constant $\Yset = \{ *\}$, the learning rate $\learningRate = 1.0$,  and 
$\timeEstimateRetro \approx \infty$.
Under this setting, we showed in Section~\ref{sec:ancestral-learning} that the update of ancestral learning is $\suptime{\pif}{i} \gets \suptime{\pib}{i-1}$, where $\suptime{\pib}{i-1}$ is the retrospective process of $\suptime{\pif}{i-1}$.
This update is equivalent to Eq.~\eqnref{eq:model-fisher-PD-fraction} if we identify $\suptime{p}{t}$ with $\suptime{\pif}{i}$.
In addition, the gain of the population fitness by evolutionary process is equivalent to the acceleration of the evolutionary process by ancestral learning. 
To see this, we introduce a measure of the acceleration defined by
$\Delta \suptime{\lambda}{i} := \lambda(\suptime{\pif}{i}) - \lambda(\suptime{\pif}{i-1})$,  where $\suptime{\pif}{i-1}$ and $\suptime{\pif}{i}$ are the strategy of the agent before and after the update by ancestral learning.
The gain $\Delta \suptime{\lambda}{i}$ of the population fitness depends on ancestral learning and is independent of natural selection.
We can therefore regard $\Delta \suptime{\lambda}{i}$ as a measure of the acceleration.
The gain $\Delta \suptime{\lambda}{i}$ is equivalent to the left-hand-side of Eq.~\eqnref{eq:FF-thm-growth-rate} if we identify $\suptime{p}{t}$ with $\suptime{\pif}{i}$ as before.
Owing to these two equivalences, we can extend the FF-thm (Eq.~\eqnref{eq:FF-thm-growth-rate}) for the population fitness to ancestral learning by substituting $\suptime{p}{t}$ with $\suptime{\pif}{i}$:
\begin{align}
    \label{thm:ff-thm-ancestral-learning}
    \Delta \suptime{\lambda}{i} &:= \lambda(\suptime{\pif}{i}) - \lambda(\suptime{\pif}{i-1})
    =  \logV_{\suptime{\pif}{i-1}}[e^{k(x)}]
\end{align}
This theorem reveals the relationship between the gain of the population fitness by an update of ancestral learning and the log-variance of the individual fitness of the strategy.

The theorem also reveals the trade-off between the acceleration $\Delta \suptime{\lambda}{i}$ and the population fitness $\lambda(\suptime{\pif}{i})$ by showing  that the acceleration is larger when the agent expresses a variety of types.
This is interepreted that the agent can obtain information about which type is fitted to the environment the best by expressing a variety of types.
We call such a situation exploratory.
On the other hand, an agent with the optimal strategy always expresses the same type under this setting (Eq.~\eqnref{eq:optimal-strategy-const-env}).
Therefore, the theorem implies that the acceleration is almost zero when the strategy is close to the optimal and $\lambda(\suptime{\pif}{i})$ is larger.
We call such a situation exploitative.
Thus, we can see the so-called exploration-exploitation trade-off in this setting.

We can further extend the FF-thm for ancestral learning to the case where the environment is not constant:
\begin{align}
    \Delta \suptime{\lambda}{i} &= \average{
        \logCov{e^{k(x,y)}}{e^{k(x,y')}}{\suptime{\pif}{i-1}}
    }{Q(y)Q(y')} \nonumber \\
    &\quad \quad + \KL{Q(y)Q(y')}{\suptime{\bar{Q}}{i}(y' \mid y) Q(y)} ,
    \label{eq:decompose-non-const-env-ffthm}
\end{align}
where
\begin{align}
    \suptime{\bar{Q}}{i}(y' \mid y)
    :\propto  \sum_{x \in \Xset} e^{k(x, y)} \suptime{\pib}{i-1}(x \mid y') Q(y') .
\end{align}
See Appendix~\ref{subsec:ffthm-ancestral-learning-deriv} for the proof.

Notice that the above equation is reduced to Eq.~\eqref{thm:ff-thm-ancestral-learning} if the environment is constant.
Eq.~\eqnref{eq:decompose-non-const-env-ffthm} is different from the FF-thm of natural selection in non-constant environment because the time evolution of $\suptime{p}{t}$ is different from the update $\suptime{\pif}{i} \gets \suptime{\pibm}{i-1}$ by ancestral learning.
The time evolution of $\suptime{p}{t}$ in the non-constant environment is stochastic and governed by
\begin{align}
    \suptime{p}{t}(x) = \frac{e^{k(x, y)}\suptime{p}{t-1}(x)}{\sum_{x' \in \Xset} e^{k(x', y)}\suptime{p}{t-1}(x')},
\end{align}
with probability $Q(y)$.

\section{Measures to characterize Ancestral Learning}
By using terms that appears in Eq.~\eqref{eq:decompose-non-const-env-ffthm}, we can quantitatively characterize different aspects of strategies during and after learning.
We define actual gain $\actualGain{i}$ and expected gain $\expectedGain{i}$ by the left and right hand sides of Eq.~\eqnref{eq:decompose-non-const-env-ffthm}, respectively:
\begin{align}
    \label{eq:actual-gain}
    \actualGain{i} := \lambda(\suptime{\pif}{i}) - \lambda (\suptime{\pif}{i-1}),
\end{align}
and 
\begin{align}
    \label{eq:expected-gain}
    \expectedGain{i} &:= \varianceTerm{i} + \KLterm{i}.
\end{align}
where $\varianceTerm{i}$ and $\KLterm{i}$ are a variance and KL terms of $\expectedGain{i}$ defined respectively as 
\begin{align}
    \label{eq:variance-term}
    \varianceTerm{i} := \average{
        \logCov{e^{k(x,y)}}{e^{k(x,y')}}{\suptime{\pif}{i-1}}
    }{Q(y)Q(y')},
\end{align}
and 
\begin{align}
    \label{eq:KL-term}
    \KLterm{i} := \KL{Q(y)Q(y')}{\suptime{\bar{Q}}{i}(y' \mid y) Q(y)}.
\end{align}

The reason why the additional KL term (Eq.~\eqnref{eq:KL-term}) appears in Eq~\eqnref{eq:decompose-non-const-env-ffthm} is attributed to the existence of two representative strategies:  \emph{bet-concentrating} and \emph{bet-balancing}.
Each term (Eqs.~\eqnref{eq:variance-term}~and~\eqnref{eq:KL-term}) of the expected gain (Eq.~\eqnref{eq:expected-gain}) is associated with one of the representative strategy and equals to the gain of the population fitness by the corresponding strategy.
Bet-concentrating is defined as a situation where an agent expresses a small subset of types that are fitted to the environment.
Formally, a strategy is  bet-concentrating on $\Xset' \subsetneq \Xset$ if $\pif(x) > 0$ for $x \in \Xset'$ and $\pif(x) = 0$ otherwise.
An example is the optimal strategy (Eq.~\eqnref{eq:optimal-strategy-const-env}) for the constant environment,
which is concentrating on the single optimal type $\{x^*\}$.
The bet-concentrating strategy is beneficial when the environment is constant or the environmental states $y \in \Yset$ are similar to each other since an agent can survive by expressing not all but a few types in such situations.
Here, similarity between two environmental states $y$ and $y'$ means the closeness of $e^{k(x,y)}$ and $e^{k(x,y')}$ for all $x \in \Xset$ (See the next paragraph for the formal definition).
However, if the environmental states are dissimilar, an agent cannot reproduce efficiently by concentrating on only a few types because those types are not adaptive to some environmental states.
An agent should stochastically choose types from a variety of alternatives to reduce the risk of bet-concentrating.
The probability to expresse a type should be determined such that the strategy has a greater population fitness.
Even if the strategy is bet-concentrating on a subset $\Xset'$ with $\# \Xset' > 1$, the probability $\pif(x)$ for $x \in \Xset'$ should be determined to maximize $\lambda$.
We define bet-balancing in $\Xset'$ as the stochastic expression of the types in $\Xset'$ whose probabilities are positive and are set so that the population fitness is maximized.
In general, the optimal strategy is the combination of bet-concentrating and bet-balancing. 
For example, let us examine the optimal strategy $\pif^* = (0.72, 0.0, 0.28)$ in the model shown in Fig.~\ref{fig:ff-thm} (j), which is calculated numerically.
The strategy is bet-concentrating on $\Xset' = \{0,2\}$ and bet-balancing in $\Xset'$.

During the evolutionary process with learning, an agent attains the optimal strategy by acquiring the two representative strategies.
The variance  and KL terms of the expected gain, $\varianceTerm{i}$ and $\KLterm{i}$,  correspond to the gains of population fitness by acquiring the respective strategy. 
The variance term $\varianceTerm{i}$ measures the gain of population fitness by acquiring bet-concentrating whereas the KL term $\KLterm{i}$ does by acquiring bet-balancing.
To see this interpretation, we rewrite the updated strategy $\suptime{\pif}{i}$.
We proved that $\suptime{\pif}{i} = \suptime{\pibm}{i-1}$ in Section~\ref{sec:ancestral-learning} when $\timeEstimateRetro \approx \infty$.
By definition,
\begin{align}
    \label{eq:retro-average-factor}
    \suptime{\pibm}{i-1}(x) &= \average{\frac{e^{k(x,y)}}{\average{e^{k(x',y)}}{\suptime{\pif}{i-1}(x')}} }{Q(y)} \suptime{\pif}{i-1}(x)\\
    &\propto \average{e^{k(x,y)}}{Q(y)}\suptime{\pif}{i-1}(x).
\end{align}
This equation is the transformation of the probability distribution $\suptime{\pif}{i-1}$ into $\suptime{\pibm}{i-1}$ by multiplying $\average{e^{k(x,y)}}{Q(y)}$ for each $x \in \Xset$.
In the transformation, the normalization factor is $\average{e^{k(x',y)}}{\suptime{\pif}{i-1}(x')Q(y)}$.
Let us examine the multiplicative factors.
For convenience, we define a vector $(\average{e^{k(x,y)}}{Q(y)})_{x \in \Xset} \in \R^{\Xset}$ by collecting the multiplicative factors for $x \in \Xset$.
It is the average of the vectors $\bm{F}_y := (e^{k(x,y)})_{x \in \Xset} \in \R^{\Xset}$ defined for each $y$.
We regard $\bm{F}_y$ as a representation of environmental state $y$ by embedding it into $\R^{\Xset}$ (Fig.~\ref{fig:ff-thm} (e,h)). 
We can use the embedding to measure similarity between the environmental states $y$ and $y'$ by  $\logCov{\fitness{y}{x}}{\fitness{y'}{x}}{\suptime{\pif}{i}}$.
By taking the normalization factor $\average{e^{k(x',y)}}{\suptime{\pif}{i-1}(x')Q(y)}$ into account, we also define a scaled embedding $\bm{f}_y$ by
\begin{align}
    \scaledFitness{y}{x} := \frac{\fitness{y}{x}}{\average{e^{k(x',y)}}{\suptime{\pif}{i-1}(x')}},
\end{align}
which depends on the current strategy $\suptime{\pif}{i-1}$ in addition to $y$.
We use the scaled embedding to rewrite Eq.~\eqnref{eq:retro-average-factor} as
\begin{align}
    \suptime{\pib}{i-1}(x) = \average{{f}_y(x)}{Q(y)}\suptime{\pif}{i-1}(x).
\end{align}
The updated strategy $\suptime{\pibm}{i-1}$ is more bet-concentrating when the environmental states are more similar
since if each $\bm{f}_y$ has similar peaks (larger components), so is their average $\suptime{\pibm}{i-1}$ (Fig.~\ref{fig:ff-thm} (e)).
Iteration of such update leads to the concentration on the types where peaks lie on.
We will see that the variance term (Eq.~\eqnref{eq:variance-term}) measures the similarity of the environmental states and corresponds to the gain of the population fitness by being bet-balancing.
On the other hand, 
$\suptime{\pibm}{i-1}$ is more bet-balancing when the environmental states are more dissimilar since if each $\bm{f}_y$ has different peaks, then their average $\suptime{\pibm}{i-1}$ becomes flat (Fig.~\ref{fig:ff-thm} (h)).
Iteration of such update leads to bet-balancing because no concentration occurs and the probabilities of expressing types are balanced so that the population fitness increases.
We will see that the KL term $\KLterm{i}$ measures the dissimilarity of the vectors and corresponds to the gain of the population fitness by being bet-balancing.
By using the correspondence, we can interpret the vanish of the KL term when the environment is constant as the unnecessity of bet-balancing.

We rewrite Eq.~\eqnref{eq:decompose-non-const-env-ffthm} to see that the variance and KL terms, $\varianceTerm{i}$ and $\KLterm{i}$,  measure the similarity and the dissimilarity of the environmental states, respectively.
We first see that the variance term measures the similarity between the environmental states.
The variance term equals
\begin{align}
    \varianceTerm{i} = &\average{ \logCov{\fitness{y}{x}} {\fitness{y'}{x}} {\suptime{\pif}{i-1}} }
        {Q(y)Q(y')}.
\end{align}
Since the log-covariance measures the similarity between two environmental states, 
the variance term measures that between all environmental states.
We can say the opposite for the KL term.
The KL term equals to 
\begin{align}
    \label{eq:ff-thm-kl-rewrite}
    \KLterm{i} = \average{ - \log \frac{ \average{\fitness{y}{x} \fitness{y'}{x} }{\suptime{\pif}{i-1}} }
            {\average{\fitness{y}{x}}{\suptime{\pif}{i}} \average{\fitness{y'}{x}}{\suptime{\pif}{i-1}}  }
        }{Q(y)Q(y')}.
\end{align}
See Appendix~\ref{subsec:ff-thm-derivation} for the proof.
The KL term is in principle larger when the environmental states are more dissimilar since the second moment $ \average{\fitness{y}{x} \fitness{y'}{x} }{\suptime{\pif}{i-1}}$ appears in the numerator, although $\average{\fitness{y}{x}}{\suptime{\pif}{i}}$ in the denominator may change the relationship.
Therefore, the KL term measures the dissimilarity of the environmental states.

\section{Numerical Validation of the FF-thm for Ancestral Learning }
\label{sec:simulation-ff-thm}

We numerically verify the FF-thm for ancestral learning.
We simulate four different models whose stochastic property $Q(y)$ of environments are different.
In each model, we investigate whether the FF-thm holds, i.e., $\actualGain{i} \approx \expectedGain{i}$.
The learning rate $\alpha = 1.0$ unless otherwise specified.
Also, we set $\timeEstimateRetro = 1000$ to avoid the fluctuation of $\retroEst$ (cf. Eq.~\eqnref{eq:ff-thm-finite-time}).

\begin{figure*}[t]
  \begin{center}
        \includegraphics[width= 0.8\linewidth]{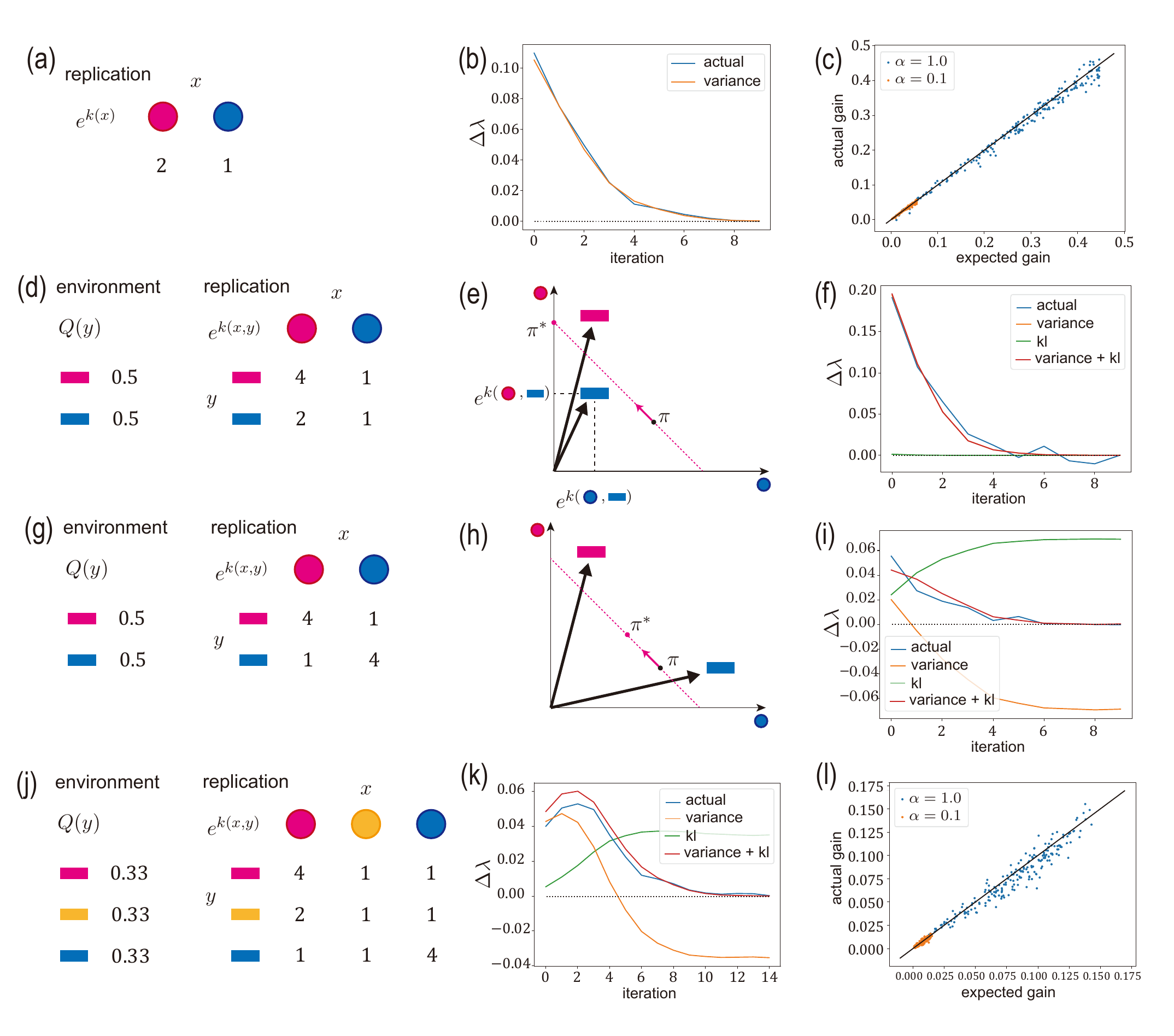}
   \end{center}
    \caption{
    The numerical validation of the FF-thm for ancestral learning. The learning rate is $\alpha = 1.0$ unless otherwise specified.  (a--c) The constant environment model (a).  The comparison between the actual gain $\actualGain{i}$ and the variance term $\varianceTerm{i}$  (Eqs.~\eqnref{eq:actual-gain} and ~\eqnref{eq:variance-term}) along the lineage of an agent (b). Notice that $\varianceTerm{i} = \expectedGain{i}$ (Eq~\eqnref{eq:expected-gain}) when the environment is constant. At each update, we observe $\actualGain{i} \approx \varianceTerm{i}$.
    The dotted black line represents $\Delta \lambda = 0$.
    The comparison between $\actualGain{1}$ and $\varianceTerm{1}$ when an agent has a randomly generated initial strategy (c).
    For most of the strategies, $\actualGain{1} \approx \varianceTerm{1}$ is observed when the learning rate is $\learningRate = 1.0$ or $\learningRate = 0.1$.
    (d--f) The similar environment model (d). An illustration of the representation of environmental state $y$ in $\R^{\Xset}$ by the embedding vector $\bm{F}_y$ (e).  The environmental state $y$ is represented so that the $x$-th component of $\bm{F}_y$ is $e^{k(x,y)}$. 
    The environmental states are similar since these two embedded vectors point to similar directions.
    The optimal solution $\pi^*$ in this model is bet-concentrating. Geometrically, the strategy lies on the red dotted line $\sum_{x \in \Xset} \pif(x) = 1$ and the optimal is on the axis corresponding to the red type. The strategy thus moves toward $\pi^*$ on this line. 
    The comparison between $\actualGain{i}$, $\varianceTerm{i}$, $\KLterm{i}$ (Eq.~\eqnref{eq:KL-term}), and $\expectedGain{i}$ (f).
    Since the FF-thm for non-constant environments (Eq.~\eqnref{eq:decompose-non-const-env-ffthm}) has the additional KL-term, the KL term $\KLterm{i}$ and the expected gain $\expectedGain{i}$ are shown in addition to (b).
    We can observe that the FF-thm holds and the variance term dominates.
    (g--i) The dissimilar environment model (g).
    An illustration of the embeddings of the environmental states in this model (h). From this embedding, we can see that the environmental states are dissimilar. In this case, the optimal strategy is bet-balancing and lies at the middle of the red dotted line. 
    The same comparison as (f) in this model (g).
    We observe that the FF-thm holds and the KL term is not negligible. 
    At last, $\varianceTerm{i} \approx \KLterm{i}$ is achieved.
    (j--l) The combined model (j). 
    Since the variance term drops earlier than the KL term, we can see that an agent learns bet-concentrating first and then acquires bet-balancing.
    The same comparison as (c) in this model (l).
    For most of the strategies, we can see that $\actualGain{1} \approx \expectedGain{1}$ holds when the learning rate is $\alpha = 1.0$ or $\alpha = 0.1$.
    }
    \label{fig:ff-thm}
\end{figure*}

We first validate the FF-thm when the environment is constant.
We simulate the model shown in Fig.~\ref{fig:ff-thm} (a) and call it a constant environment model.
We observe that $\actualGain{i} \approx \expectedGain{i}$ along the lineage of an agent whose initial strategy is $\suptime{\pif}{0} = (0.5, 0.5)$  (Fig.~\ref{fig:ff-thm} (b)).
To check the validity of the FF-thm beyond one lineage, we compare $\actualGain{1}$ and $\expectedGain{1}$ of the agent that has an initial strategy generated uniformly at random (Fig.~\ref{fig:ff-thm} (c)).
We observe that $\actualGain{1} \approx \expectedGain{1}$ for most of the random strategies.

We next verify the FF-thm when the environment is not constant by simulating three models.
We first simulate the model shown in (Fig.~\ref{fig:ff-thm} (d)).
Since the environmental states are similar in this model, we call it a similar environment  model.
In this model, the optimal strategy is bet-concentrating (Fig.~\ref{fig:ff-thm} (e)) on $\{0\}$ and the variance term $\varianceTerm{i}$  is expected to dominate.
Fig.~\ref{fig:ff-thm} (f) shows $\actualGain{i}$, $\expectedGain{i}$, $\varianceTerm{i}$, and $\KLterm{i}$ along the lineage of an agent whose initial strategy is $\suptime{\pif}{0} = (0.5, 0.5)$.
From the plot, we find that $\actualGain{i} \approx \expectedGain{i}$ and that the variance term dominates as expected.

We next simulate the model shown in Fig.~\ref{fig:ff-thm} (g).
Since environmental states are dissimilar in this model, we call it a dissimilar environment model.
In this model, the optimal strategy is bet-balancing as illustrated in Fig.~\ref{fig:ff-thm} (h), and the KL term (Eq.~\eqnref{eq:KL-term}) is expected to be non-negligible.
Fig.~\ref{fig:ff-thm} (i) shows $\actualGain{i}$, $\expectedGain{i}$, $\varianceTerm{i}$, and $\KLterm{i}$ along the lineage of an agent whose initial strategy is $\suptime{\pif}{0} = (0.9, 0.1)$.
We verify $\actualGain{i} \approx \expectedGain{i}$ and find that the KL term is not negligible as expected.
We also observe that $\varianceTerm{i} \approx \KLterm{i}$ as $i$ increases.

We finally simulated the model shown in Fig.~\ref{fig:ff-thm} (j).
In this model, the environmental state $0$ and $1$ are similar whereas the state $2$ is dissimilar from them.
Therefore, we call the model a combined model.
In this model, the optimal strategy $\pi^* = (0.72, 0, 0.28)$ is the combination of bet-concentrating on $\Xset' = \{0, 2\}$ and bet-balancing over $\Xset'$.
Fig.~\ref{fig:ff-thm} (k) shows $\actualGain{i}$, $\expectedGain{i}$, $\varianceTerm{i}$, and $\KLterm{i}$ along the lineage of an agent whose initial strategy is $(0.05,0.15, 0.8)$. We can see that $\actualGain{i} \approx \expectedGain{i}$.
We also observe that the KL term is not negligible.
Since the variance term drops faster than the KL term, an agent acquires the bet-concentrating strategy first and then does the bet-balancing strategy.
This interpretation is also supported from the strategy $\suptime{\pif}{5} = (0.31, 0.04, 0.65)$ just before the fifth update, when the variance term becomes negative for the first time.
The strategy is almost concentrating on $\Xset' = \{0,2\}$.
On the other hand, the strategy is not bet-balancing in $\Xset'$ since $\suptime{\pif}{5}(0)$ and $\suptime{\pif}{5}(2)$ are far from the optimal probabilities $\pif^*(0)$ and $\pif^*(2)$, respectively.
To check the validity of the FF-thm beyond one lineage, we compare $\actualGain{1}$ and $\expectedGain{1}$ of the agent that has an initial strategy generated uniformly at random (Fig.~\ref{fig:ff-thm} (l)).
We observe that $\actualGain{1} \approx \expectedGain{1}$ for most of the random strategies.

\section{Tradeoff between learning rate and update interval}
\label{sec:tradeoff}

The FF-thm for ancestral learning is derived for $\learningRate =1$ and $\timeEstimateRetro\gg 1$. 
To address other situations, especially one where $\timeEstimateRetro$ is not so large, we further extend the FF-thm for ancestral learning to the case where learning rate $\learningRate < 1.0$ and show that there is a trade-off relation between $\learningRate$ and $\timeEstimateRetro$.
First, we define an $\learningRate$-log-covariance by generalizing Eq.~\eqnref{eq:covariance-and-log-covariance}:
\begin{align}
    \logCovAlpha{f(x)}{g(x)}{p} := \log \left(\learningRate \frac{\cov{f(x)}{g(x)}{p}}{\average{f(x)}{p}\average{g(x)}{p}} + 1\right).
\end{align}
By using this quantity, we have
\begin{align}
    &\Delta \suptime{\lambda}{i} \nonumber\\
    &= \average{\logCovAlpha{e^{k(x,y)}}{e^{k(x,y')}}{\suptime{\pif}{i-1}}}{Q(y)Q(y')} \nonumber \\
    &\quad\quad + \KL{Q(y)Q(y')}{\suptime{\bar{Q}}{i}_{\learningRate}(y' \mid y) Q(y)},
    \label{eq:ff-thm-non-const-env-learning-rate}
\end{align}
where
\begin{align}
    \suptime{\bar{Q}}{i}_\alpha(y' \mid y)
    :\propto  \sum_{x \in \Xset} e^{k(x, y)} \suptime{\pibAlpha}{i-1}(x \mid y') Q(y'),
\end{align}
and
\begin{align}
    \suptime{\pibAlpha}{i-1}(x \mid y') = \learningRate \suptime{\pib}{i-1}(x \mid y') + (1 - \learningRate) \suptime{\pif}{i-1}(x) .
\end{align}
See Appendix~\ref{subsec:ff-thm-learning-rate} for the proof.
We again define the actual and expected gains, which generalize Eqs.~\eqnref{eq:actual-gain} and~\eqnref{eq:expected-gain}, by the left and right hand sides of Eq.~\eqnref{eq:ff-thm-non-const-env-learning-rate} respectively as
\begin{align}
    \label{eq:actual-gain-learning-rate}
    \actualGain{i} := \lambda(\suptime{\pif}{i}) - \lambda (\suptime{\pif}{i-1}),    
\end{align}
and
\begin{align}
    \label{eq:expected-gain-learning-rate}
    \expectedGain{i} := & \average{\logCovAlpha{e^{k(x,y)}}{e^{k(x,y')}}{\suptime{\pif}{i-1}}}{Q(y)Q(y')} \nonumber \\
    &\quad\quad + \KL{Q(y)Q(y')}{\suptime{\bar{Q}}{i}_{\learningRate}(y' \mid y) Q(y)}.
\end{align}

To check the validity of the FF-thm (Eq.~\eqnref{eq:expected-gain-learning-rate} for $\learningRate < 1.0$,
we simulate the constant environment model (Fig.~\ref{fig:ancestral_learning} (a)) and the combined model (Fig.~\ref{fig:ff-thm} (j)) when the learning rate is $\learningRate = 0.1$.
We compare $\actualGain{1}$ and $\expectedGain{1}$ of the agent that has an initial strategy generated uniformly at random (Fig.~\ref{fig:ff-thm} (c,l)).
We observe that $\actualGain{1} \approx \expectedGain{1}$ for most of the random strategies.

When $\timeEstimateRetro < \infty$, the FF-thm (Eq.~\eqnref{eq:ff-thm-non-const-env-learning-rate}) does not hold and  $\actualGain{i} < \expectedGain{i}$. 
Owing to the finite update interval, the ancestral information $\retroEst^{\suptime{\pif}{i-1}}$ and the updated strategy $\suptime{\pif}{i} = \learningRate \retroEst^{\suptime{\pif}{i-1}} + (1 - \alpha) \suptime{\pif}{i-1}$ fluctuate around their expectation $\suptime{\pibm}{i-1}$ and $\suptime{\pibAlpha}{i-1}$, respectively.
The averaged population fitness $\average{\lambda(\suptime{\pif}{i})}{}$ with respect to this fluctuation is smaller than $\lambda(\suptime{\pibAlpha}{i-1})$ by the concavity of $\lambda$ and the Jeansen's inequality.
When $\timeEstimateRetro$ is sufficiently large (but still finite), we can quantify this decrease by
\begin{align}
    \label{eq:ff-thm-finite-time}
    \actualGain{i} \approx \expectedGain{i} + \frac{\learningRate^2}{2} \trace{\bm{I}_\lambda\bm{V} }.
\end{align}

Here, $\trace{\bm{A}}$ is the trace of matrix $\bm{A}$, the matrix $\bm{V}$ is the covariance matrix of $\retroEst$ defined by
\begin{align}
    V(x,x') = \average{\retroEst(x) \retroEst(x')}{} - \pibm(x)\pibm(x'),
\end{align}
and
\begin{align}
    \fim(x,x') = \frac{\partial^2 \lambda(\pibm)}{\partial \pi(x) \partial \pi(x')}.
\end{align}
See Appendix~\ref{subsec:ff-thm-finite-time} for the proof.
We note that the second term is non-positive due to the negative semidefiniteness of $\bm{I}_\lambda$ shown from the concavity of $\lambda$.
Since $\bm{V}$ is of the order $1 / \timeEstimateRetro$, the deviation $\learningRate^2  \trace{\bm{I}_\lambda\bm{V} } / 2$  from the FF-thm for $\timeEstimateRetro = \infty$  is negligible if the learning rate is sufficiently small compared to update interval $\timeEstimateRetro$: $\learningRate^2 / \timeEstimateRetro \ll 1$.
Thus, there is a trade-off between $\learningRate$ and $\timeEstimateRetro$ in relation with the efficiency of learning.

In Section~\ref{sec:ff-thm}, we mainly focused on the case of $\timeEstimateRetro = \infty$ to make the FF-thm (Eq.~\eqnref{eq:ff-thm-non-const-env-learning-rate}) intuitive.
However, a short $\timeEstimateRetro$ is realistic and might be beneficial in both biological and engineering systems. 
The benefit of a short $\timeEstimateRetro$ is that an agent has more opportunities for the acceleration by the update of strategy.
The drawback is that the acceleration by each update becomes smaller due to the fluctuation of $\retroEst$  around its expectation $\pibm$.
Equation ~\eqnref{eq:ff-thm-finite-time} indicates that the decrease is of the order $\learningRate^2 / \timeEstimateRetro$.
It implies that an agent can keep the decrease small by adopting small $\learningRate$ compared to $\timeEstimateRetro$, although such a small learning rate makes the learning slow (Eq.~\eqnref{eq:ff-thm-non-const-env-learning-rate}).
In other words, the decrease in memory size $\timeEstimateRetro$ can be compensated by the decrease in learning speed $\learningRate$.
Since the decrease of the acceleration (Eq.~\eqnref{eq:ff-thm-finite-time}) depends on the second power of $\learningRate$ while it does on the first power of $\timeEstimateRetro$, an agent might prefer the pair of small $\learningRate$ and short $\timeEstimateRetro$ to  that of large $\learningRate$ and long $\timeEstimateRetro$.
Indeed, we have numerically shown that ancestral learning accelerates the evolutionary process with small $\learningRate = 0.01$ and short $\timeEstimateRetro = 1$ in Section~\ref{sec:ancestral-learning}.
In such a situation, our extended FF-thm is insightful because the deviation (Eq.~\eqnref{eq:ff-thm-finite-time}) is small.

\section{Discussion}
\label{sec:discussion}

In the present paper, we investigated the acceleration of the evolutionary process by learning.
We first numerically showed that ancestral learning can accelerates the evolutionary process.
We next proved that an agent can estimate the gradient of the population fitness from the ancestral information $\retroEst$ without the communication between agents.
We  then quantified the acceleration via extending the FF-thm for the ancestral learning and revealed that the gain of the population fitness by ancestral learning has a connection to the log-variance of the individual fitness of the strategy.
We finally derived the trade-off relation between the learning rate and the update interval.
Overall, we have established a theoretical framework to characterize and evaluate the impacts of learning in evolutionary processes. 

However, there remain some sorts of factors that might be useful for agents to learn but we have not considered.
 One is the type of a parent.
While an agent with ancestral learning uses the ancestor's types $\retroEst$, it does not use the type of the parent directly.
Such strong dependence on the parent might be beneficial when the environmental state is strongly correlated to the previous state.
When type $x$  of an agent depends on that $x'$ of the parent, the type-switching strategy should be modeled as a Markov transition $\tranF(x \mid x')$ instead of the distribution $\pif(x)$.
Promising techniques for the generalization are the large deviation and the variational representation, which played the important role in the present paper, for Markov chains in random environments~\cite{seppalainen1994large, kifer1996perron}.

Another  one is communications between agents.
Although we showed that the agent can estimate the gradient without communications, learning with such information might further accelerate the evolutionary process than ancestral learning.
The acceleration by ancestral learning becomes small when the update interval $\timeEstimateRetro$ is short due to the fluctuation of $\retroEst$ (Eq.~\eqnref{eq:ff-thm-finite-time}).
Communications between agents might be useful to suppress such fluctuation.

The last one is sensing of the environmental state.
In the context of population dynamics, researchers have considered the situation where an agent receives a sensing signal $z$ of the environmental state $y$ and then expresses their type by a signal-dependent strategy $\pif(x \mid z)$~\cite{haccou1995optimal, kussell2005phenotypic, rivoire2011value, kobayashi2015fluctuation}.
Since sensing is another form of information processing, we should consider the unification of sensing and learning to understand the significance of information processing to organisms.
In such a setting, an agent might attain the optimal strategy $\pif^*(x \mid z)$ via extended ancestral learning.
Also, such an sensing signal might improve ancestral learning.
To achieve such unification, we need a theory that can integrate the prospective and retrospective information obtained by sensing and learning.

\section*{Acknoledgements}
The first author is supported by JSPS Research Fellowship Grant Number JP19J22607 and JST ACT-X Grant Number JPMJAX190L.
This research is supported by JSPS KAKENHI Grant Numbers 19H05799 and 19H03216 and by JST CREST JPMJCR1927 and JPMJCR2011.

\section*{Source code availability}
The source code for simulation is available at \url{https://github.com/so-nakashima/learning_in_growing_systems}.
The language was C++17 with Boost 107100.
We used Windows Subsystem for Linux 2.
The operating system was Ubuntu 20.04.1 LTS on Windows 10 version 2004.
We used gcc 9.3.0 for compiling.
For Fig.~\ref{fig:lineage-tree-learning}, we used graphviz 2.43.0 (0) and colormap (\url{https://github.com/jgreitemann/colormap}).
For the other plottings, we used matplotlib-cpp (\url{https://github.com/lava/matplotlib-cpp}), which requires Python 3.
We used Python 3.8.5.


\section{Derivations and proof}
\label{sec:derivations}

\subsection{Variational representation of the growth rate (Eq.~\eqref{eq:variation-growth-rate})}
\label{subsec:variational-rep-derivation}
The proof is a special case of~\cite{sughiyama2015pathwise, kobayashi2015fluctuation}.
For the completeness of the paper, we give the proof.
For a fixed $y \in \Yset$ and an arbitrary distribution $\pi$ over $\Xset$,
\begin{align}
    \log \average{e^{k(x,y)}}{\pif(x)} &= \log \sum_{x \in \Xset} \pi(x) \frac{\pif(x)}{\pi(x)} e^{k(x,y)}
\end{align}
By applying the Jensen's inequality, we have
\begin{align}
    \log \average{e^{k(x,y)}}{\pif(x)} &\ge \sum_{x \in \Xset} \pi(x) \left[\log \frac{\pif(x)}{\pi(x)} e^{k(x,y)}\right]\\
    &= \sum_{x \in \Xset} \pi(x) \left[k(x,y) - \log \frac{\pi(x)}{\pif(x)} \right]\\
    &= \sum_{x \in \Xset} \pi(x) k(x,y) - \KL{\pi}{\pif}.
\end{align}
By substituting $\pi(x)$ with $\pib(x \mid y)$, we can see that the equality is attained.
Therefore, 
\begin{align}
    \label{eq:variation-before-expectation}
    \log \average{e^{k(x,y)}}{\pif(x)} = \max_{\pi}\left\{\sum_{x \in \Xset} k(x)\pi(x) - \KL{\pi}{\pif} \right\}.
\end{align}
By averaging the equality with respect to $Q(y)$, we have Eq.~\eqref{eq:variation-growth-rate}.

\subsection{Gradient of the growth rate (Eqs.~\eqnref{eq:derivative-lambda}~and ~\eqnref{eq:gradient-pib})}
\label{subsec:gradient-growth-rate}

The proof is essentially the same as~\cite{sughiyama2015pathwise}.
Since the maximizer of the right hand side of Eq.~\eqnref{eq:variation-before-expectation} is $\pib(x \mid y)$,
\begin{align}
     \log \average{e^{k(x,y)}}{\pif(x)} = \sum_{x \in \Xset} k(x, y) \pib(x \mid y) - \KL{\pib}{\pif}.
\end{align}
We differentiate the both hand sides with respect to $\pif(x)$ while taking into account of the dependence of $\pib$ on $\pif$:
\begin{align}
    &\frac{\partial}{\partial \pif(x)}  \log \average{e^{k(x,y)}}{\pif(x)}\\
    &= \frac{\partial  \KL{\pib}{\pif}}{\partial \pif(x)} + \sum_{x' \in \Xset} \frac{\partial \pib(x' \mid y)}{\partial \pif(x)} \frac{\partial F[\pib]}{\partial \pib(x' \mid y)},
\end{align}
where $F[\pi] := \sum_{x \in \Xset} k(x, y)\pi(x) - \KL{\pi}{\pif} $.
Since $\pib$ is the maximizer of the $F$, the derivative of $F$ at $\pib$ is zero and consequently the seconde term vanishes.
Therefore,
\begin{align}
     \frac{\partial}{\partial \pif(x)}\log \average{e^{k(x,y)}}{\pif(x)} = \frac{\pib(x \mid y)}{\pif(x)}.
\end{align}
By taking average with respect to $Q(y)$, we have Eq.~\eqnref{eq:derivative-lambda}.

We next prove Eq.~\eqnref{eq:gradient-pib} via the method of Lagrange multiplier.
For sufficiently small $\epsilon$, we need to solve the following linearized optimization:
\begin{align}
    \max_{\delta \pi}. \sum_{x \in \Xset} \frac{\pibm(x)}{\pif(x)} \delta \pi(x)
\end{align}
under the constraints $\sum_{x \in \Xset} \delta \pi(x) = 0$ and $\KL{\pif}{\pif + \delta \pi} = \epsilon$.
For a sufficiently small $\epsilon$, we can approximate $\KL{\pif}{\pif + \delta \pi}$ by using the \emph{Fisher information matrix}~\cite{amari2016information} as
\begin{align}
    \KL{\pif}{\pif + \delta \pi} &= \frac{1}{2}\sum_{x,x' \in \Xset} \delta\pi(x)\delta_{x,x'} \frac{1}{\pif(x)} \delta\pi(x')\\
    &= \frac{1}{2}\sum_{x\in \Xset}  \frac{\delta\pi^2(x)}{\pif(x)}.
\end{align}
Here, the Fisher information matrix is a $|\Xset| \times |\Xset|$ diagonal matrix with diagonal entries $\{1 / \pif(x) \}_{x \in \Xset}$.
By using this approximation, the Lagrangian function is
\begin{align}
    L(\delta \pi ; \lambda, \lambda') &= \sum_{x \in \Xset} \frac{\pibm(x)}{\pif(x)} \delta \pi(x) \nonumber \\
        &\quad\quad + \frac{\lambda}{2} \left(\sum_{x \in \Xset} \frac{\delta\pi^2(x)}{\pif(x)}  - \epsilon \right) \nonumber \\
        &\quad\quad + \lambda' \left(\sum_{x \in \Xset} \delta \pi(x) \right).
\end{align}
By differentiating $L$ with respect to $\delta \pi(x)$, we have the stationary condition:
\begin{align}
    \label{eq:lagrangian-deriv-growth-rate}
    \frac{\partial L}{\partial \delta \pi(x)} = \frac{\pibm(x)}{\pif(x)} + \frac{\lambda \delta \pi(x)}{\pif(x)} + \lambda' = 0, 
\end{align}
for all $x \in \Xset$.
By multiplying $\pif(x)$ and taking sum $\sum_{x \in \Xset}$ of the both hand side of Eq.~\eqnref{eq:lagrangian-deriv-growth-rate}, we have
\begin{align}
    1 + \lambda' = 0.
\end{align}
We here used $\sum_{x \in \Xset} \delta \pi (x) = 0$.
By rearranging Eq.~\eqnref{eq:lagrangian-deriv-growth-rate} and substituting $\lambda' = -1$, we have
\begin{align}
    \delta \pi (x) = \frac{\pif(x) - \pibm(x)}{\lambda} \propto \pibm(x) - \pif(x).
\end{align}

\subsection{Fisher's fundamental theorem of natural selection (Eqs. \eqref{eq:FF-thm-mean-fitness} and \eqref{eq:FF-thm-growth-rate})}
\label{subsec:ff-thm-derivation}
We first prove Eq.~\eqref{eq:FF-thm-mean-fitness} for the completeness of the paper.
By direct calculation,
\begin{align}
     &\Delta \average{e^{k(x)}}{\suptime{p}{t}}\\
     &= \sum_{x \in \Xset} e^{k(x)} \suptime{p}{t}(x) - \sum_{x \in \Xset} e^{k(x)} \suptime{p}{t-1}(x)\\
     &=  \sum_{x \in \Xset} e^{k(x)} \frac{e^{k(x)}\suptime{p}{t-1}(x)}{\sum_{x' \in \Xset }e^{k(x')}\suptime{p}{t-1}(x')} \nonumber\\
     &\quad\quad - \sum_{x \in \Xset} e^{k(x)} \suptime{p}{t-1}(x)\\
     &= \frac{\sum_{x \in \Xset} \left(e^{k(x)}\right)^2 \suptime{p}{t-1}(x) - \left(\sum_{x \in \Xset} e^{k(x)} \suptime{p}{t-1}(x) \right)^2}{\sum_{x' \in \Xset }e^{k(x')}\suptime{p}{t-1}(x')}\\
     &= \frac{\V_{\suptime{p}{t-1}}\left[e^{k(x)} \right]}{ \average{e^{k(x)}}{\suptime{p}{t-1}}}.
\end{align}

We next prove Eq.~\eqref{eq:FF-thm-growth-rate}.
By direct calculation,
\begin{align}
    &\Delta \suptime{\lambda}{t}\\
    &= \log \sum_{x \in \Xset}e^{k(x)} \suptime{p}{t}(x) - \log \sum_{x \in \Xset} e^{k(x)} \suptime{p}{t-1}(x)\\
    &= \log \sum_{x \in \Xset} e^{k(x)} \frac{e^{k(x)}\suptime{p}{t-1}(x)}{\sum_{x' \in \Xset }e^{k(x')}\suptime{p}{t-1}(x')} \nonumber\\
    &\quad \quad  - \log \sum_{x \in \Xset} e^{k(x)} \suptime{p}{t-1}(x)\\
    &= \log \sum_{x \in \Xset} \left( e^{k(x)}\right)^2 \suptime{p}{t-1}(x)
     - 2 \log \sum_{x \in \Xset} e^{k(x)} \suptime{p}{t-1}(x)\\
    &=\log\frac{ \average{\left( e^{k(x)}\right)^2}{\suptime{p}{t-1}} } { \average{e^{k(x)}}{\suptime{p}{t-1}}^2 }\\
    &= \logV_{\suptime{p}{t-1}}[e^{k(x)}].
\end{align}

\subsection{Fisher's fundamental theorem of ancestral learning for non-constant environment (Eqs.~\eqnref{eq:decompose-non-const-env-ffthm} and~\eqnref{eq:ff-thm-kl-rewrite})}
\label{subsec:ffthm-ancestral-learning-deriv}
We first prove Eq.~\eqnref{eq:decompose-non-const-env-ffthm}.
By direct calculation,
\begin{align}
    & \lambda(\suptime{\pif}{i})\\
    &= \average{\log \average{e^{k(x,y)}}{\suptime{\pibm}{i-1}}  }{Q(y)}\\
    &= \average{\log \average{e^{k(x,y)}}{\suptime{\pibm}{i-1}} }{Q(y)Q(y')}\\
    &= \average{ \log \average{e^{k(x,y)}}{\suptime{\pib}{i-1}(x \mid y')}  }{Q(y)Q(y')} \nonumber\\
    &\quad \quad + \average{ \log \frac{ \average{e^{k(x,y)}}{\suptime{\pibm}{i-1}} }{  \average{e^{k(x,y)}}{\suptime{\pib}{i-1}(x \mid y')}}  }{ Q(y) Q(y')}.
    \label{eq:ffthm-nonconst-env-deriv-intermediate}
\end{align}
We first treat the first term.
By a similar argument to Eq.~\eqnref{eq:FF-thm-growth-rate}, the term inside the expectation satisfies the following relationship.
\begin{align}
     &\log \average{e^{k(x,y)}}{\suptime{\pib}{i-1}(x \mid y')} -\log \average{ e^{k(x,y)}}{ \suptime{\pif}{i-1}}\\
    &= \log \sum_{x \in \Xset} \frac{e^{k(x,y)+k(x,y')} \suptime{\pif}{i-1}(x)}{\average{e^{k(x',y')}}{\suptime{\pif}{i-1}(x')}} \nonumber\\
    &\quad\quad -  \log \average{ e^{k(x,y)}}{ \suptime{\pif}{i-1}}\\
    &= \log \frac{\average{e^{k(x,y) + k(x,y')}}{\suptime{\pif}{i-1}}}{\average{e^{k(x,y)}}{\suptime{\pif}{i-1}} \average{e^{k(x,y')}}{\suptime{\pif}{i-1}}  }
\end{align}
By taking average with respect to $Q(y)Q(y')$, we have
\begin{align}
    &\average{ \log \average{e^{k(x,y)}}{\suptime{\pib}{i-1} (x \mid y')}  }{Q(y)Q(y')} - \lambda(\suptime{\pif}{i-1})\\
    &= \average{\log \frac{\average{e^{k(x,y) + k(x,y')}}{\suptime{\pif}{i-1}}}{ \average{e^{k(x,y)}}{\suptime{\pif}{i-1}} \average{e^{k(x,y')}}{\suptime{\pif}{i-1}}   }}{Q(y)Q(y')}\\
    &= \average{\logCov{e^{k(x,y)}}{e^{k(x,y')}}{\suptime{\pif}{i-1}}}{Q(y)Q(y')}.
\end{align}

We next treat the second term of Eq.~\eqnref{eq:ffthm-nonconst-env-deriv-intermediate}.
By definition,
\begin{align}
    \label{eq:derive-ff-thm-KL}
     &\frac{\suptime{\bar{Q}}{i-1}(y' \mid y) }{ Q(y')}\\
     &= \frac{\sum_{x \in \Xset} e^{k(x,y)} \suptime{\pib}{i-1}(x \mid y')}{\sum_{x \in \Xset, y' \in \Yset}  e^{k(x,y)} \suptime{\pib}{i-1}(x \mid y') Q(y') }\\
     &= \frac{\average{e^{k(x,y)}}{\suptime{\pib}{i-1}(x \mid y')}}{\average{e^{k(x,y)}}{\suptime{\pibm}{i-1}}}
\end{align}
We thus have
\begin{align}
    &\average{ \log \frac{ \average{e^{k(x,y)}}{\suptime{\pibm}{i-1}} }{  \average{e^{k(x,y)}}{\suptime{\pib}{i-1}(x \mid y') }}  }{ Q(y) Q(y')}\\
    &= \average{ \log \frac{ Q(y')  }{ \suptime{\bar{Q}}{i-1}(y' \mid y) } }{ Q(y)Q(y')}\\
    &=  \average{ \log \frac{ Q(y)Q(y')  }{ \suptime{\bar{Q}}{i-1}(y' \mid y) Q(y)} }{ Q(y) Q(y')}\\
    &= \KL{Q(y)Q(y')}{ \suptime{\bar{Q}}{i-1}(y' \mid y) Q(y)}.
\end{align}
In conclusion, we proved Eq.~\eqnref{eq:decompose-non-const-env-ffthm}.

We next prove Eq.~\eqnref{eq:ff-thm-kl-rewrite}.
By Eq.~\eqnref{eq:derive-ff-thm-KL},
\begin{align}
    & \log \frac{Q(y)Q(y')}{ \suptime{\bar{Q}}{i}(y' \mid y) Q(y)  }\\
    &= - \log \frac{\average{e^{k(x,y)}}{\suptime{\pib}{i-1}(x \mid y')}}{\average{e^{k(x,y)}}{\suptime{\pibm}{i-1}}}\\
    &= - \log \frac{\average{e^{k(x,y) + k(x, y')}}{\suptime{\pif}{i-1}}}
        {\average{e^{k(x,y)}}{\suptime{\pibm}{i-1}} \average{e^{k(x,y')}}{\suptime{\pif}{i-1}}}\\
    &= - \log \frac{ \average{\fitness{y}{x} \fitness{y'}{x} }{\suptime{\pif}{i-1}} }
            {\average{\fitness{y}{x}}{\suptime{\pif}{i}} \average{\fitness{y'}{x}}{\suptime{\pif}{i-1}}  }.
\end{align}
By averaging with respect to $Q(y)Q(y')$, we have Eq.~\eqnref{eq:ff-thm-kl-rewrite}.

\subsection{Fisher's fundamental theorem of ancestral learning when $\learningRate < 1$ (Eq.~\eqnref{eq:ff-thm-non-const-env-learning-rate})}
\label{subsec:ff-thm-learning-rate}
We can prove Eq.~\eqnref{eq:ff-thm-non-const-env-learning-rate} by almost the same argument as Eq.~\eqnref{eq:decompose-non-const-env-ffthm}.
Let $\suptime{\pibAlphaM}{i-1}(x) := \learningRate \suptime{\pibm}{i-1}(x) + (1 - \alpha) \suptime{\pif}{i-1}(x)$.
By direct calculation, we have
\begin{align}
    & \lambda(\suptime{\pif}{i})\\
    &= \average{\log \average{e^{k(x,y)}}{\suptime{\pibAlphaM}{i-1}}  }{Q(y)}\\
    &= \average{\log \average{e^{k(x,y)}}{\suptime{\pibAlphaM}{i-1}} }{Q(y)Q(y')}\\
    &= \average{ \log \average{e^{k(x,y)}}{\suptime{\pibAlpha}{i-1}(x \mid y')}  }{Q(y)Q(y')} \nonumber\\
    &\quad \quad + \average{ \log \frac{ \average{e^{k(x,y)}}{\suptime{\pibAlphaM}{i-1}} }{  \average{e^{k(x,y)}}{\suptime{\pibAlpha}{i-1}(x \mid y')} }  }{ Q(y) Q(y')}.
    \label{eq:ffthm-learning-rate-deriv-intermediate}
\end{align}
We first treat the first term.
By a similar argument to Eq.~\eqnref{eq:FF-thm-growth-rate}, the term inside the expectation satisfies
\begin{align}
     &\log \average{e^{k(x,y)}}{\suptime{\pibAlpha}{i-1}(x \mid y')} -\log \average{ e^{k(x,y)}}{ \suptime{\pif}{i-1}}\\
    &= \log \average{ 
        e^{k(x,y)} \left(\alpha \frac{e^{k(x,y')}}{ \average{e^{k(x',y')}}{\suptime{\pif}{i-1}(x')}  } + 1 - \alpha \right)}{\suptime{\pif}{i-1}(x)} 
        \nonumber\\
    &\quad\quad -  \log \average{ e^{k(x,y)}}{ \suptime{\pif}{i-1}}\\
    &= \log \left(\alpha \frac{\average{e^{k(x,y) + k(x,y')}}{\suptime{\pif}{i-1} } }{ \average{e^{k(x,y')}}{\suptime{\pif}{i-1}}}   + (1 - \alpha) \average{e^{k(x,y)}}{\suptime{\pif}{i-1}} \right) 
    &\quad\quad -  \log \average{ e^{k(x,y)}}{ \suptime{\pif}{i-1}}\\
    &= \log \left( \alpha \frac{ \average{e^{k(x,y) + k(x,y')}}{\suptime{\pif}{i-1}}}
        { \average{e^{k(x,y)}}{\suptime{\pif}{i-1}} \average{e^{k(x,y')}}{\suptime{\pif}{i-1}}    } + 1 - \alpha \right)
\end{align}
By similar argument to Eq.~\eqnref{eq:covariance-and-log-covariance},
\begin{align}
     &\log \average{e^{k(x,y)}}{\suptime{\pibAlpha}{i-1}(x \mid y')} -\log \average{ e^{k(x,y)}}{ \suptime{\pif}{i-1}}\\
     &=  \log \left( \alpha \frac{ \cov{\fitness{y}{x}}{\fitness{y'}{x}}{\suptime{\pif}{i-1}} }
        { \average{\fitness{y}{x}}{\suptime{\pif}{i-1}} \average{\fitness{y'}{x}}{\suptime{\pif}{i-1}}    } + 1 \right)\\
    &= \logCovAlpha{e^{k(x,y)}}{e^{k(x,y')}}{\suptime{\pif}{i-1}}.
\end{align}
By taking average with respect to $Q(y)Q(y')$, we have
\begin{align}
    &\average{ \log \average{e^{k(x,y)}}{\suptime{\pibAlpha}{i-1}(x \mid y')}  }{Q(y)Q(y')} - \lambda(\suptime{\pif}{i-1})\\
    &=   \average{
        \logCovAlpha{e^{k(x,y)}}{e^{k(x,y')}}{\suptime{\pif}{i-1}}
    }{Q(y)Q(y')}.
\end{align}

We next treat the second term of Eq.~\eqnref{eq:ffthm-nonconst-env-deriv-intermediate}.
By definition,
\begin{align}
     &\frac{\suptime{\bar{Q}}{i-1}_{\learningRate}(y' \mid y) }{ Q(y')}\\
     &= \frac{\sum_{x \in \Xset} e^{k(x,y)} \suptime{\pibAlpha}{i-1}(x \mid y')}{\sum_{x \in \Xset, y' \in \Yset}  e^{k(x,y)} \suptime{\pibAlpha}{i-1}(x \mid y') Q(y') }\\
     &= \frac{\average{e^{k(x,y)}}{\suptime{\pibAlpha}{i-1}(x \mid y')}}{\average{e^{k(x,y)}}{\suptime{\pibAlphaM}{i-1}}}
\end{align}
Thus,
\begin{align}
    &\average{ \log \frac{ \average{e^{k(x,y)}}{\suptime{\pibAlphaM}{i-1}} }{  \average{e^{k(x,y)}}{\suptime{\pibAlpha}{i-1}(x \mid y')}}  }{ Q(y) Q(y')}\\
    &= \average{ \log \frac{ Q(y')  }{ \suptime{\bar{Q}}{i-1}_{\learningRate}(y' \mid y) } }{ Q(y)Q(y')}\\
    &=  \average{ \log \frac{ Q(y)Q(y')  }{ \suptime{\bar{Q}}{i-1}_{\learningRate}(y' \mid y) Q(y)} }{ Q(y) Q(y')}\\
    &= \KL{Q(y)Q(y')}{ \suptime{\bar{Q}}{i-1}_{\learningRate}(y' \mid y) Q(y)}.
\end{align}
In conclusion, we proved Eq.~\eqnref{eq:ff-thm-non-const-env-learning-rate}.

\subsection{Fisher's fundamental theorem of ancestral learning when $\timeEstimateRetro$ is finite (Eq.~\eqnref{eq:ff-thm-finite-time})}
\label{subsec:ff-thm-finite-time}

When $\timeEstimateRetro$ is sufficiently large (but finite), we can approximate $\retroEst$ by the central limit theorem~\cite{vaart1998asymptotic} as
\begin{align}
     \retroEst \sim \normal{\suptime{\pib}{i-1}}{\bm{V}},
\end{align}
where $\normal{\bm{\mu}}{\bm{\Sigma}}$ is the multivariate normal distribution with mean $\bm{\mu}$ and covariance $\bm{\Sigma}$.
The updated strategy $\suptime{\pif}{i} = \alpha \retroEst + (1 - \alpha) \suptime{\pif}{i-1}$ satisfies
\begin{align}
    \suptime{\pif}{i} \sim \normal{\pi_\learningRate}{\alpha^2 \bm{V}},
\end{align}
where we omitte the superscript of $\suptime{\pibAlphaM}{i-1}$ to avoid the complication.
The growth rate is approximated as
\begin{align}
    \lambda(\pi_\learningRate + \delta \pi) &\approx \lambda(\pi_\learningRate) + \sum_{x \in \Xset} \frac{\partial \lambda}{\partial \pi(x)} \delta \pi(x) \nonumber\\
    &\quad \quad + \frac{1}{2} \sum_{x,x' \in \Xset} \delta \pi(x) \frac{\partial^2 \lambda}{\partial \pi(x) \pi(x')} \delta \pi(x')\\
    &=\lambda(\pi_\alpha) + \sum_{x \in \Xset} \frac{\partial \lambda}{\partial \pi(x)} \delta \pi(x) \nonumber\\
    &\quad \quad + \frac{1}{2} \sum_{x,x' \in \Xset} \delta \pi(x) \fim(x,x') \delta \pi(x')\\
\end{align}
By this approximation,
\begin{align}
    \actualGain{i} &= \average{\lambda(\retroEst)}{} - \suptime{\lambda}{i-1}\\
    &\approx \lambda(\pi_\learningRate) - \suptime{\lambda}{i-1} + \average{\sum_{x \in \Xset} \frac{\partial \lambda}{\partial \pi(x)} \delta \pi(x)}{\normal{0}{\alpha^2 \bm{V}}} \nonumber\\
    &\quad\quad + \frac{1}{2}\average{\sum_{x,x' \in \Xset} \delta \pi(x) \fim(x,x') \delta \pi(x')}{\normal{0}{\alpha^2 \bm{V}}}\\
    &= \expectedGain{i}  + \frac{1}{2} \average{\sum_{x,x' \in \Xset} \delta \pi(x) \fim(x,x') \delta \pi(x')}{\normal{0}{\alpha^2 \bm{V}}}.
\end{align}
In the last equation, the third term vanishes because
\begin{align}
    &\average{\sum_{x \in \Xset} \frac{\partial \lambda}{\partial \pi(x)} \delta \pi(x)}{\normal{0}{\alpha^2 \bm{V}}}\\
    &= \sum_{x \in \Xset} \frac{\partial \lambda}{\partial \pi(x)} \average{\delta \pi(x) }{\normal{0}{\alpha^2 \bm{V}}}\\
    &= 0.
\end{align}
By the usual matrix calculation~\cite{petersen2012matrix},
\begin{align}
    & \average{\sum_{x,x'} \delta \pi(x) \fim(x,x') \delta \pi(x')}{\normal{0}{\alpha^2 \bm{V}}}\\
    &= \alpha^2 \trace{\bm{I}_\lambda \bm{V}}.
\end{align}
In all, we proved \eqnref{eq:ff-thm-finite-time}.


\bibliography{main}

\end{document}